\documentclass[onecolumn,showpacs,superscriptaddress]{revtex4}
\usepackage{amsmath}
\usepackage{amssymb}

\usepackage{graphicx}
\graphicspath{{img/}}

\renewcommand{\vec}[1]{{\mathbf{#1}}}
\newcommand{\mat}[1]{{\mathsf{#1}}}
\newcommand{\Tr}{\operatorname{Tr}}

\newcommand{\T}{{\cal T}}
\newcommand{\A}{{\cal A}}
\newcommand{\B}{{\mat{B}}}
\newcommand{\D}{{\mat{D}}}
\newcommand{\xs}{{x_{s}}}
\newcommand{\xu}{{x_{u}}}
\newcommand{\Ord}{{\cal O}}
\newcommand{\xbs}{\bar x_s}
\newcommand{\xbu}{\bar x_u}
\newcommand{\xbsu}{\bar x_{s,u}}
\newcommand{\PA}{P_{\cal A}}
\newcommand{\opt}{{\rm opt}}
\newcommand{\bnabla}{{\boldsymbol \nabla}}
\newcommand{\DF}{{\bnabla\vec{F}(\vec{x}^{\ast}_{k}(t),t)}}
\renewcommand{\d}{{\mathrm{d}}}

\begin{document}
\title{Activated escape over oscillating barriers: The case of many 
dimensions}
\author{J\"org Lehmann}
\affiliation{Institut f\"ur Physik, Universit\"at Augsburg,
             Universit\"atsstra\ss e~1, D-86135 Augsburg, Germany}
\author{Peter Reimann}
\affiliation{Fakult\"at f\"ur Physik, Universit\"at Bielefeld,
             Universit\"atsstra\ss e~25, D-33615 Bielefeld, Germany}
\author{Peter H\"anggi}
\affiliation{Institut f\"ur Physik, Universit\"at Augsburg,
             Universit\"atsstra\ss e~1, D-86135 Augsburg, Germany}
\date{\today}
\begin{abstract}
  We present a novel path-integral method for the determination of time-dependent
  and time-averaged reaction
  rates in {\em multidimensional}, periodically driven escape problems at weak thermal noise.
   The so obtained general expressions are evaluated
  explicitly for the situation of a sinusoidally driven, damped particle with inertia moving in a
  metastable, piecewise parabolic potential.
  A comparison with data from Monte-Carlo simulations yields a very good
  agreement with our analytic results over a wide parameter range.
\end{abstract}
\maketitle

\section{Introduction}
\label{introduction}

Thermally activated escape problems in the presence of an explicit
time-dependent driving are at the root of many timely transport
processes. Typical examples comprise the control of chemical reactions
with tailored laser pulses \cite{control01,control00}, ion transport
through voltage-gated channels \cite{ser}, the pumping and shuttling
of particles in Brownian environments
\cite{rat1,rat2,rat3,rei02,reiph02,ASTPH}, or the amplification of
weak information-carrying signals via the phenomenon of Stochastic
Resonance \cite{jun91,sr,Moss,CHEMPHYSCHEM}, to name only a few. In
the absence of such a time-dependent driving and for the case of weak
thermal noise the escape time is governed -- as commonly known -- by
an exponentially leading Arrhenius factor \cite{han90,fle93,tal95}.
Pioneered by Kramers \cite{kra40} and extended to arbitrary dimensions
in the works \cite{lan61,lan69,mat77,tal87}, this scheme, however,
meets formidable difficulties under far from thermal equilibrium
conditions. This is so because of an extremely complex interplay
between the {\em global} properties of the metastable potential and
the nonlinear noisy dynamics \cite{han90,gra85,ein95,mai97}.  The
subject of our present paper is one of the simplest and experimentally
most natural such non-equilibrium descendants of Kramers' original
escape problem \cite{kra40}, namely the thermally activated escape of
a Brownian particle over a potential barrier in the presence of
periodic driving which modulates both the corresponding potential well
region and the activation barrier. While most previous attempts have
been restricted to weak \cite{gra84,jun93,sme99}, slow
\cite{jun89,tal99}, or fast \cite{gra84,jun89,rei96a} driving, we have
addressed in recent analytical explorations \cite{leh00a,leh00b} by
means of a path-integral technique the most challenging intermediate
regime of {\it moderately strong} and {\em moderately fast} driving
for a one-dimensional, overdamped escape problem.  Closely related to
our recent works are the subsequent appealing efforts in Ref.
\cite{mai01}, wherein on uses instead the method of singular
perturbation theory in the weak noise limit --- a so termed WKB
approximation --- which, however, also has been restricted to cover only
the overdamped, one-dimensional case.  The scheme in
Ref.~\cite{mai01} yields results which are consistent with the
findings in Ref.~\cite{leh00a,leh00b}; but it remains on a more formal
and implicit level compared to ours.

The objective of this study is to extend these recent works of
time-dependent rate theory put forward in
Refs.~\cite{leh00a,leh00b,mai01} to the case of many dimension of the
underlying stochastic process. In particular, we shall consider the
generic case of a driven inertial Brownian motion dynamics.
Furthermore, we shall derive for weak noise asymptotically exact
results in analytically closed form for the escape of an inertial,
sinusoidally driven Brownian particle in a metastable, piecewise
parabolic potential in the regime of moderate forcing strengths and
forcing frequencies. A comparison with data from Monte-Carlo
simulations yields very good agreement over a wide parameter range.

We anticipate here that the following presentation is on purpose
kept rather concise. While we are confident that the general idea
of our approach remains more transparent in this way, it is
nevertheless advisable to consult the detailed discussion of the
one-dimensional, overdamped (i.e. no inertial) dynamics in
Ref.~\cite{leh00b} for a more thorough and in-depth understanding.
It must be pointed out, however, that our present generalization
to many dimensions in addition requires several conceptually new
steps and features as compared to the one-dimensional overdamped
case.

In Sect.~\ref{escape}, we describe the model under investigation and
define both the instantaneous and the time-averaged escape rate. The
path-integral formalism and its evaluation for weak-noise is described
in Sect.~\ref{pathintegral}.  Its application to the escape problem is
demonstrated in Sect.~\ref{rateformula}. Our general findings are then
evaluated and discussed explicitly for the case of the driven Kramers
problem with a piecewise parabolic potential in Sect.~\ref{example}.
Concluding remarks are given in Sect.~\ref{conclusion}.

\section{The escape problem}
\label{escape}

\subsection{The general Model}

The starting-point of our investigations is the following model for the
$d$-dimensional Brownian motion of a particle with coordinates $\vec{x}(t)$
in a time-dependent force field $\vec{F}(\vec{x},t)$:
\begin{equation}
  \dot{\vec{x}}(t) = \vec{F}(\vec{x}(t),t) + \sqrt{2\, \epsilon}\,\B\,
  \boldsymbol \xi (t)
  \ .
  \label{2.1}
\end{equation}
Here, bold quantities denote $d$-dimensional vectors, while
$d\times d$-matrices are represented by sans-serif fonts. The
$d$-dimensional Gaussian uncorrelated, white noise ${\boldsymbol
\xi(t)}$ is defined by the relations ($i,j=1,2,\dots,d$)
\begin{equation}
  \langle\xi_{i}(t)\rangle = 0
  \ , \quad
  \langle\xi_i(t)\,\xi_j(t')\rangle = \delta_{ij}\,\delta (t-t') \ ,
  \label{2.2}
\end{equation}
where $\delta_{ij}$ is the Kronecker delta and $\delta(t)$ Dirac's
$\delta$-distribution. The additive coupling of the noise to the stochastic
dynamics is determined by the matrix~$\B$, which we assume to be non-singular.

We restrict ourselves to the case of a time-periodic driving, \textit{i.e.}\
$\vec{F}(\vec{x},t)=\vec{F}(\vec{x},t+\T)$ for a certain period $\T$.
Furthermore, we assume that the deterministic dynamics, \textit{i.e.}\
Eq.~(\ref{2.1}) for $\epsilon=0$, possesses exactly one stable periodic
orbit~$\vec{x}_{s}(t)=\vec{x}_{s}(t+\T)$ with a time-dependent domain of
attraction $\A(t)$. All other deterministic orbits, which start outside of
$\A(t)$ and its boundary ${\cal S}(t):=\partial \A(t)$ are assumed to diverge in the long
time limit.  The boundary ${\cal S}(t)$ then acts as a
separatrix between these two kinds of deterministic solutions. Moreover, we
require that there is exactly one unstable periodic orbit
$\vec{x}_{u}(t)=\vec{x}_{u}(t+\T)$, which then moves with the separatrix,
\textit{i.e.}\ $\vec{x}_u(t)\in S(t)$ for all times $t$.  Finally, all other
deterministic solutions starting in a neighborhood of the unstable periodic
orbit on the separatrix are assumed to be bounded.
All these assumptions can usually be taken for granted
in typical periodically driven escape problems of practical interest.
\begin{figure}
  \begin{center}
    \includegraphics[width=0.45\linewidth]{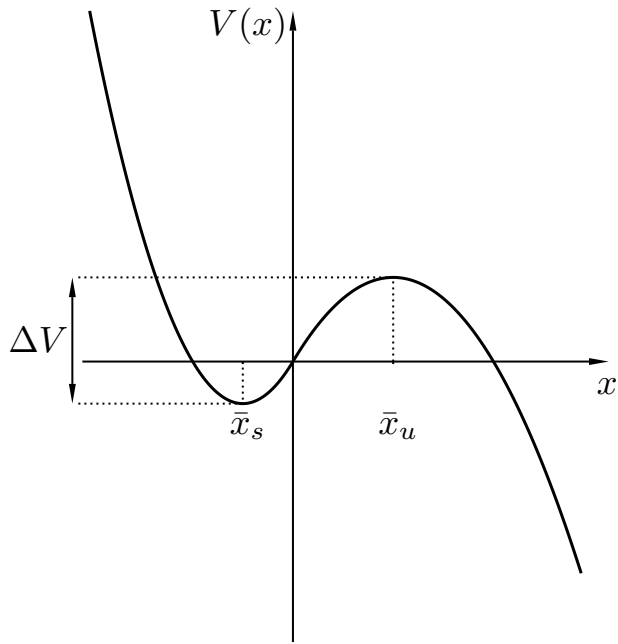}
    \caption{Sketch of a typical metastable potential $V(x)$, namely
      the piecewise parabolic potential~\eqref{8.1} with $\Delta
      V =0.9$, $m\,\omega_s^2= 0.6$, and $m\,\omega_u^2=-0.3$ in
      arbitrary, dimensionless units.}
    \label{fig:potential}
  \end{center}
\end{figure}

\begin{figure}[htbp]
  \includegraphics[width=0.5\linewidth]{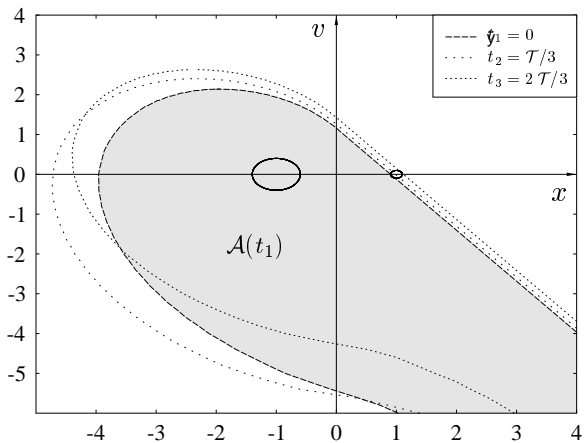}
  \caption{Phase space diagram for a typical metastable potential with
    additive sinusoidal driving~\eqref{2.3}, namely the piecewise parabolic
    potential \eqref{8.1} with $\xbs = \omega_u^2 = -1$, $\xbu =
    \omega_s^2 = m = \Delta V = \Omega = 1$, $\eta=0.5$, $A=0.2$. Solid line:
    Stable (left) and unstable (right) periodic orbit. Dashed lines:
    Separatrix at different times $t_1=0$, $t_2=\T/3$, and $t_3=2\T/3$. Grey
    area: Domain of attraction ${\cal A}(t_1)$ of stable periodic orbit
    $x_s(t)$ at time $t_1$.}
  \label{fig:separatrix}
\end{figure}

A particularly prominent example is the Kramers problem
\cite{kra40}, namely the escape of a particle with mass $m$ out of
the bottom well of a static potential as cartooned in
Fig.~\ref{fig:potential}.  Under the additional influence of an
additive sinusoidal driving this corresponds to the stochastic
dynamics
\begin{equation}
  m \,\ddot{x} + \eta \,\dot{x} = - V'(x) + A \sin(\Omega\,t) +
  \sqrt{2 \eta k_{B}T} \,\xi(t)\ .
\label{2.3}
\end{equation}
Here, $\eta$ is the viscous friction coefficient, $k_{B}$
Boltzmann's constant, $T$ the temperature and $\Omega=2\pi/\T$ the
angular frequency of the driving.  We can rewrite the second order
equation~\eqref{2.3} in the form of Eq.~\eqref{2.1}, if we
identify
\begin{equation}
  \vec{x}:=
  \begin{pmatrix}
    x\\v
  \end{pmatrix}
  \ ,\quad
  \vec{F}(\vec{x},t) :=
  \begin{pmatrix}
    v\\\frac{1}{m}\,F(x,t) - \gamma \, v
  \end{pmatrix}
  \ ,\quad
  \B:=
  \frac{\sqrt{\eta}}{m}
  \begin{pmatrix}
    \delta & 0 \\
    0 & 1
  \end{pmatrix}
  \ , \quad
  \epsilon:=k_{B}\,T
  \ ,
  \label{2.4}
\end{equation}
where we have introduced the one-dimensional force field $F(x,t):= - V'(x) + A
\sin(\Omega\,t)$ and the frequency $\gamma:=\eta/m$. Furthermore, we
have added an auxiliary noise source of a strength proportional to $\delta$
to ensure that $\B$ is non-singular. Eventually, we shall consider the
limit $\delta\to0$, in which Eq.~\eqref{2.4} reduces to the original
dynamics~\eqref{2.3}.
Figure~\ref{fig:separatrix} depicts the stable and unstable periodic orbits,
the domain of attraction ${\cal A}(t)$, and the separatrix ${\cal S}(t)$ for a
representative metastable potential.

A fully equivalent description of the stochastic dynamics~\eqref{2.1} is
provided by the Fokker-Planck equation \cite{ris84} for the probability
distribution $p(\vec{x},t)$ of an ensemble of particles $\vec{x}(t)$,
\begin{equation}
  \frac{\partial}{\partial t}\, p(\vec{x},t) + \nabla \cdot \vec{j}(\vec{x},t)
  = 0 \ ,\label{2.9a}
\end{equation}
where the probability current density is given by
\begin{align}
  \vec{j}(\vec{x},t) :=
  \vec{F}(\vec{x},t)\,p(\vec{x},t)
  - \epsilon\,\mat{D}\bnabla p(\vec{x},t)
  \label{2.5}
\end{align}
with the positive semidefinite diffusion matrix $\mat{D}:= \B\B^{T}$.

\subsection{Escape rates}

If we now consider our stochastic dynamics (\ref{2.1}) for finite
noise-strengths $\epsilon$, it is well-known that a particle $x(t)$ starting
in ${\cal A}(t)$ will be able to escape out of this basin.  To quantify this
escape of particles, we first introduce the population $P_{\cal A}(t)$ in
${\cal A}(t)$:
\begin{equation}
  P_{{\cal A}}(t) := \int\limits_{{\cal A}(t)} \!\! \d^{d}\vec{x} \;p(\vec{x},t)
  \ .
  \label{2.7}
\end{equation}
A natural definition of an ``instantaneous rate'' $\Gamma (t)$ for such
escape events is then given by the relative decrease of this population per unit
time \cite{leh00a, leh00b},
\begin{equation}
\Gamma (t) := -\dot \PA (t)/\PA (t) \ .
\label{2.8}
\end{equation}
We remark that --- except for transients at early times --- this
quantity is independent of the initial conditions at time $t=t_{0}$,
provided we start with a distribution that is concentrated within the
basin ${\cal A}(t)$. We thus make the convenient choice
$p(\vec{x},t_0) = \delta (\vec{x}-\vec{x}_s(t_0))$. Furthermore, on
the time scale of these transients, $\Gamma(t)$ will approach a
time-periodic limit, permitting a meaningful definition of the
time-averaged rate
\begin{equation}
  \bar\Gamma := \frac{1}{\T}\,\int_{t}^{t+\T}\!\!\!\d t'\, \Gamma(t') \ .
  \label{2.12}
\end{equation}
In the sequel of the present paper we will deduce analytic expressions for
both the instantaneous and the time-averaged rate in the limit of small
noise-strength $\epsilon$.

To make progress in this direction, we first note that for weak noise
$\epsilon$ the typical time scale of the escape events $1/\bar\Gamma$ is well
separated from the time scale of the just mentioned transients~\cite{han90}.
We can thus approximate $P_{\cal A}(t)$ by its initial value $P_{\cal
  A}(t_{0})=1$ in the denominator of Eq.~(\ref{2.8}). Using the Fokker-Planck
equation (\ref{2.9a}) and the divergence theorem, the instantaneous rate reads
\begin{equation}
  \label{2.13}
  \Gamma(t) = \int\limits_{{\cal S}(t)} \!\!
               \d^{d-1}\vec{n}\cdot
               \left[\,
                 \vec{j}(\vec{x},t) -
                 \dot{\vec{x}} \, p(\vec{x},t)
               \right] \ .
\end{equation}
Here, the integration is over the entire time-dependent separatrix ${\cal
  S}(t)$ and $\vec{n}$ denotes their outer normal vector.
Furthermore, the time derivative in $\dot{\vec{x}}$ refers to the
$t$-dependence of the points on this separatrix, which is
determined by the deterministic equation of motion, \textit{i.e.}
Eq.~(\ref{2.1}) with $\epsilon=0$. Taking this and the definition
(\ref{2.5}) of the current density into account, one obtains the
general result
\begin{equation}
  \label{2.14}
  \Gamma(t) = - \epsilon \!\!\!\int\limits_{{\cal S}(t)}\!\!
                    \d^{d-1}\vec{n} \cdot \mat{D} \bnabla p(\vec{x},t)
                    \ .
\end{equation}
As expected on naive grounds, a crossing of the separatrix is
possible via a diffusive process only.

\section{Path integrals and the saddle-point approximation}
\label{pathintegral}

Let us briefly summarize the path integral description of the dynamics
generated by a stochastic differential equation~\eqref{2.1}.  It is well-known
from the literature
\cite{gra85,ein95,mai97,lud75,wio89,fre84,wei79,car81,sch81,lan82,han89,han93}
that one can represent the conditional probability density
$p(\vec{x_{f}},t_{f}|\vec{x}_{0},t_{0})$ to find the particle at time $t_{f}$
at position $\vec{x}_{f}$ if it has started at time $t_{0}$ at position
$\vec{x}_{0}$ as weighted sum over all paths between these two points,
\begin{equation}
  \label{3.3}
  p(\vec{x}_{f},t_{f}|\vec{x}_{0},t_{0}) =
  \!\!\int\limits_{\vec{x}(t_{0})=\vec{x}_{0}}^{\vec{x}(t_{f})=\vec{x}_{f}}\!\!
  {\cal D}\vec{x}(t)\, e^{-S[x(t)]/\epsilon}\ .
\end{equation}
Here, $S[x(t)]$ is the effective action or the Onsager-Machlup
functional, \textit{i.e.}
\begin{equation}
  \label{3.4}
  S[x(t)] := \int_{t_{0}}^{t_{f}} \! \d t
  \,L(\vec{x}(t),\dot{\vec{x}}(t),t)\ ,
\end{equation}
with the corresponding Lagrangian
\begin{equation}
  \label{3.5}
  L(\vec{x},\dot{\vec{x}},t)
  := \frac{1}{4}\left[\dot{\vec{x}} - \vec{F}(\vec{x},t)\right] 
      \cdot \mat{D}^{-1}
      \left[\dot{\vec{x}} - \vec{F}(\vec{x},t)\right]\ .
\end{equation}
We remark that a prepoint-discretization scheme
\cite{lan82,han89,han93} (not to be confused with the Ito-scheme
in stochastic calculus \cite{ris84}) has been implicitly adopted
in the path integral~(\ref{3.3}) implying the measure
$1/\sqrt{\det (4\pi\,\epsilon\, \Delta t\,\mat{D})}$. Other
``discretization schemes'' \cite{lan82,han89,han93} would give
rise to a somewhat modified path-integral formalism but would ---
of course --- lead to identical results as far as the actual
stochastic dynamics (\ref{2.3}) is concerned.

It should be clear that in general an exact evaluation of the path
integral~\eqref{3.3} is impossible. Nevertheless, it represents an
advantageous starting-point for a systematic weak-noise approximation: For
small $\epsilon$, the main contributions to the path integral~\eqref{3.3}
stem from the surrounding of action minimizing paths. Carrying out a
infinite-dimensional saddle-point approximation around these paths, one
arrives at
\begin{equation}
  \label{3.7}
  p(\vec{x}_{f},t_{f}|\vec{x}_{0},t_{0}) =
  \sum_{k}
  \frac{e^{-\phi_{k}(\vec{x}_{f},t_{f})/\epsilon}}
       {\left[
           \det ( 4\,\pi\,\epsilon\, \mat{Q}^{\ast}_{k}(t_{f}) )
        \right]^{\frac{1}{2}}
       } \,
  [1+\Ord(\epsilon)]\ ,
\end{equation}
where the sum runs over all paths $\vec{x}^{\ast}_{k}(t)$ that (locally) minimize the
action~(\ref{3.4}). These obey the Euler-Lagrange equations
\begin{equation}
  \ddot{\vec{x}}^{\ast}_{k}(t)  +
  \left[
    \D\, (\DF)^T\, \D^{-1} -
    \,\DF
  \right] \dot{\vec{x}}^\ast_k(t)
  - \D\, (\DF)^T\, \D^{-1}\, \vec{F}(\vec{x}^\ast_k(t),t)
  - \frac{\partial}{\partial t}\vec{F}(\vec{x}^{\ast}_{k}(t),t) = 0
  \label{3.7a}
\end{equation}
together with the boundary conditions
\begin{equation}
  \label{3.9}
  \vec{x}^{\ast}_{k}(t_{0}) = \vec{x}_{0}
  \quad,\quad
  \vec{x}^{\ast}_{k}(t_{f}) = \vec{x}_{f}\ .
\end{equation}
Here and in the following, $\bnabla \vec{F}(\vec{x},t)$ denotes the Jacobian
matrix with components $(\bnabla \vec{F}(\vec{x},t))_{ij}:=\partial
F_i(\vec{x},t)/\partial x_j$ and the transpose of a matrix $\mat{M}$ is
written as $\mat{M}^T$.
Introducing the canonical momentum
\begin{equation}
  \vec{p}:=\frac{\partial L}{\partial \dot{\vec{x}}} =
  \frac{1}{2} \,\mat{D}^{-1}
  \left[
    \dot{\vec{x}} - \vec{F}(\vec{x},t)
  \right]\ ,
\end{equation}
we can, via the usual Legendre transformation, pass to the equivalent
Hamiltonian dynamics, defined by the Hamiltonian
\begin{equation}
  H(\vec{x}, \vec{p}, t) := \vec{p}\cdot\dot{\vec{x}} - L =
  \vec{p}\cdot\mat{D}\,\vec{p} + \vec{p}\cdot\vec{F}(\vec{x},t)
  \label{Ham}
\end{equation}
and the corresponding canonical equations
\begin{align}
  \dot{\vec{x}}^{\ast}_{k}(t) & =
  \vec{F}(\vec{x}^{\ast}_{k}(t),t)
  + 2 \, \mat{D} \, \vec{p}^{\ast}_{k}(t)
  \nonumber\\
  \dot{\vec{p}}^{\ast}_{k}(t) & =
  -\left(\DF\right)^{T} \vec{p}^{\ast}_{k}(t)\ .
  \label{3.8}
\end{align}
The value of the action $\phi_{k}(\vec{x}_{f},t_{f})$ of a path
$\vec{x}^\ast_k(t)$ is then given by
\begin{equation}
  \label{3.10}
  \phi_{k}(\vec{x}_{f},t_{f}) := S[\vec{x}^{\ast}_{k}(t_{f})] =
  \int_{t_{0}}^{t_{f}}\!\d t\,\vec{p}^{\ast}_{k}(t)\cdot\mat{D}\,\vec{p}^{\ast}_{k}(t)\ ,
\end{equation}
where we have suppressed in favor of notational brevity the
dependence on the initial condition $\vec{x}_{0}$ at time $t_{0}$.
It is noteworthy to point out  that in Eq.~(\ref{Ham}) the
momentum  enters both quadratically and linearly: the latter
linear contribution  mimics a magnetic field-like, time-dependent
vector potential contribution. For later use, we also recall the
well-known result from classical mechanics \cite{Goldstein} that
the derivative of the extremal action with respect to its
endpoint $\vec{x}_f$ equals the canonical conjugate momentum at
time $t_f$,
\begin{equation}
  \label{3.10a}
  \vec{p}^\ast_k(t_f) =
  \frac{\partial \phi_{k}(\vec{x}_{f},t_{f})}{\partial \vec{x}_f}
  \ .
\end{equation}

The yet unspecified prefactor term $\mat{Q}^{\ast}_{k}(t_{f})$
in (\ref{3.7}) is
given by the solution at time $t_{f}$ of the following second order
differential equation
\begin{multline}
  \label{3.11}
  \ddot{\mat{Q}}^{\ast}_{k}(t) -
  \DF\,\dot{\mat{Q}}^{\ast}_{k}(t)
  - \dot{\mat{Q}}^{\ast}_{k}(t)\, \mat{Q}^{\ast}_{k}(t)^{-1}\, \mat{D}\, (\DF)^T\, \mat{D}^{-1}\,\mat{Q}^{\ast}_{k}(t) \\
  + \bigg\{
       \DF\,\mat{D}\,(\DF)^T\,\mat{D}^{-1}
     - \mat{D}\,(\DF)^T\,\mat{D}^{-1}\,\DF \\
       - \frac{\d}{\d t}\! \left[ \DF + \mat{D}\,(\DF)^T\,\mat{D}^{-1}\right]
       + 2\, (\mat{D}\, \vec{p}^{\ast}_{k}(t)) \cdot (\bnabla\bnabla\,\vec{F}(\vec{x}^{\ast}_{k}(t),t))
    \bigg\} \mat{Q}^{\ast}_{k}(t) = 0
\end{multline}
with initial conditions
\begin{equation}
  \label{3.12}
  \mat{Q}^{\ast}_{k}(t_{0}) = \mat{0} \ ,\quad \dot{\mat{Q}}^{\ast}_{k}(t_{0}) = \mat{D}\ .
\end{equation}
Here, we have introduced the tensor of third order $(\bnabla \bnabla
\vec{F}(\vec{x},t))_{ijl}:=\partial^2 F_i(\vec{x},t)/\partial x_j
\partial x_l$, whereby the scalar product with a matrix appearing in
Eq.~\eqref{3.11} is defined as
\begin{equation}
[(\mat{D} \, \vec{p}^\ast_k(t))
\cdot (\bnabla\bnabla\,\vec{F}(\vec{x}^{\ast}_{k}(t),t))]_{ij}
=
\sum_l
[\mat{D} \,\vec{p}^\ast_k(t)]_l
\frac{\partial F_l(\vec{x^\ast_k},t))}
     {\partial x_i \partial x_j}
\ .
\end{equation}
The derivation of Eqs.~(\ref{3.11},\ref{3.12}) in the framework of the
time-discretized version of the path integral~\eqref{3.3} proceeds
along the same general lines as in the one-dimensional case presented in
Ref.~\cite{leh00b}. However, the rather technical details are beyond the
scope of the present paper.

Though in principle, the prefactor $\mat{Q}^{\ast}_{k}(t_{f})$ is completely determined by
Eqs.~(\ref{3.11},\ref{3.12}), it will turn out to be more convenient to
consider the following quantity:
\begin{equation}
  \label{3.13}
  \mat{G}^{\ast}_{k}(t) :=
  \frac{1}{2}
  \bigg[
    \mat{D}^{-1}\,\dot{\mat{Q}}^{\ast}_{k}(t)\,\mat{Q}^{\ast}_{k}(t)^{-1}
    -  (\DF)^T\,\mat{D}^{-1}
     - \mat{D}^{-1}\,\DF
  \bigg]\ .
\end{equation}
With the help of Eq.~(\ref{3.11}) one can then verify that
$\mat{G}^{\ast}_{k}(t)$ fulfills for $t>t_{0}$ the matrix Riccati
equation
\begin{equation}
  \label{3.14}
  \begin{split}
  \dot{\mat{G}}^{\ast}_{k}(t) = &
  - 2\, \mat{G}^{\ast}_{k}(t)\, \mat{D}\, \mat{G}^{\ast}_{k}(t)
  - (\DF)^T\,\mat{G}^{\ast}_{k}(t) \\
  & -   \mat{G}^{\ast}_{k}(t)\,\DF
  - \vec{p}^{\ast}_{k}(t) \cdot \bnabla\bnabla\,\vec{F}(\vec{x}^{\ast}_{k}(t),t)
  \ .
  \end{split}
\end{equation}
However, as can be inferred from Eqs.~(\ref{3.12},\ref{3.13}), there is no well
defined initial condition for $\mat{G}^{\ast}_{k}(t)$ at time $t_{0}$.
Instead, we have to content ourselves with the relations
\begin{equation}
  \label{3.15}
  \lim_{t\to t_{0}} \left[\mat{G}^{\ast}_{k}(t)\, \mat{Q}^{\ast}_{k}(t)\right] = \frac{1}{2} \,
  \mat{1}
  \quad\text{and}\quad
  \mat{G}^{\ast}_{k}(t_{0})^{-1} = \mat{0}\ .
\end{equation}
By tracing over both sides of Eq.~(\ref{3.13}) one obtains a linear first
order differential equation for the determinant $\det \mat{Q}^{\ast}_{k}(t)$
from Eq.~(\ref{3.7}):
\begin{equation}
  \label{3.16}
  \frac{\d}{\d t} \det \mat{Q}^{\ast}_{k}(t) =
  2 \Tr\left[(\DF)^T + \mat{D}\, \mat{G}^{\ast}_{k}(t) \right] \det \mat{Q}^{\ast}_{k}(t)\ .
\end{equation}
The initial condition $\det\mat{Q}^{\ast}_{k}(t_{0})=0$ follows immediately
from Eq.~(\ref{3.12}).

We conclude this section with two remarks on the above mentioned two different approaches for
the calculation of the prefactor in Eq.~(\ref{3.7}).  First, we want to point
out that one can identify $\mat{G}^{\ast}_{k}(t_{f})$ with the Hessian of the
action function $\phi_k(\vec{x}_{f},t_{f})$:
\begin{equation}
  \label{3.17}
  \mat{G}^{\ast}_{k}(t_{f}) =
  \frac{\partial^{2} \phi_{k}(\vec{x}_{f},t_{f})}{\partial \vec{x}_{f}\,\partial \vec{x}_{f}}\ .
\end{equation}
The proof of this relation proceeds analogously to that presented in
Ref. \cite{leh00b} for the one-dimensional case, though the
calculational details are by far more involved. Thus, Eq.~(\ref{3.14})
is equivalent to the matrix Riccati equation for the second
derivatives of the action used elsewhere in the
literature~\cite{mai97}. However, while there this equation is derived
by inserting a WKB-type ansatz in the Fokker-Planck
equation~(\ref{2.9a}), we entirely work here within the path integral
formalism.  In particular, we avoid problems commonly encountered due
to the non-analytic nature of the WKB-action near the separatrix
\cite{gra85,mai97}. As a second remark, we note that an advantage of
Eqs.~(\ref{3.14}, \ref{3.16}) over Eq.~(\ref{3.11}) lies in the fact
that in the former set of equations no $\mat{D}^{-1}$ terms appear.
Since later on we will be interested in the case of singular
$\mat{D}$, this fact presents a favorable feature from a technical
point of view.

\section{Evaluation of the escape rate for weak noise}
\label{rateformula}

For small noise strengths $\epsilon$, we can now insert the saddle-point
approximation \eqref{3.7} into the rate expression \eqref{2.14}. Using
Eq.~\eqref{3.10a}, this yields
\begin{equation}
  \label{4.1}
  \Gamma(t_f) =
  \!\!\int\limits_{{\cal S}(t_f)}\!\!\!
  \d^{d-1}\vec{n} \cdot
  \mat{D}
  \sum_k
  \vec{p}^\ast_k(t_f)
  \frac{e^{-\phi_{k}(\vec{x}^\ast_{k}(t_f),t_{f})/\epsilon}}
       {\left[
           \det ( 4\,\pi\,\epsilon\, \mat{Q}^{\ast}_{k}(t_{f}) )
        \right]^{\frac{1}{2}}
       }
  [1+\Ord(\epsilon)]\ .
\end{equation}
Note that for the integrand the boundary condition \eqref{3.9} with
$\vec{x}_f$ being the integration variable on the separatrix ${\cal S}(t_f)$
is implicitly understood. Parametrizing the surface integral~\eqref{4.1} by
the $d-1$ dimensional vector $\vec{s}=(s_1, \dots, s_{d-1})^T$, \textit{i.e.}\
${\cal S}(t_f) =: \{ \vec{x}_\mathrm{sep}(\vec{s}, t_f)\vert\vec{s} \in
U^{d-1}\}$ for a certain subset $U^{d-1}\subseteq\mathbb{R}^{d-1}$ of
the $d-1$ dimensional parameter space, we obtain
\begin{equation}
  \label{4.2}
  \Gamma(t_f) =
  \!\!\int
  \d^{d-1}s \,
  \sqrt{g(\vec{s},t_f)}\,
  \vec{n}(\vec{s},t_f) \cdot
  \mat{D}
  \sum_k
  \vec{p}^\ast_k(t_f)
  \frac{e^{-\phi_{k}(\vec{x}^\ast_{k}(t_f),t_{f})/\epsilon}}
       {\left[
           \det ( 4\,\pi\,\epsilon\, \mat{Q}^{\ast}_{k}(t_{f}) )
        \right]^{\frac{1}{2}}
       }
  [1+\Ord(\epsilon)]\ ,
\end{equation}
where we have introduced the measure
\begin{equation}
  \label{4.3}
  g(\vec{s}, t_f) := \det
  \left(
    \frac{\partial \vec{x}_\mathrm{sep}(\vec{s},t_f)}{\partial s_\nu}
    \cdot
    \frac{\partial \vec{x}_\mathrm{sep}(\vec{s},t_f)}{\partial s_{\nu'}}
  \right)_{\nu,\nu'=1,\dots d-1}\ .
\end{equation}

For weak noise, the main contribution to the integral~\eqref{4.2}
comes again from the surrounding of the minima of the actions
$\phi_{k}(\vec{x}^\ast_{k}(t_f),t_{f})$, and we can evaluate the
integral in a saddle-point approximation. The corresponding
extremal condition which determines the values $\vec{s}^\ast_k$
then assumes the form, cf. also Eq.~\eqref{3.10a},
\begin{equation}
  \label{4.4}
  \frac{\partial \phi_k(\vec{x}_\mathrm{sep}(\vec{s}^\ast_k,t_f),t_f)}
       {\partial s^\ast_{k,\nu}} =
  \vec{p}^\ast_k(t_f) \cdot
  \frac{\partial \vec{x}_\mathrm{sep}(\vec{s}^\ast_k,t_f)}
       {\partial s^\ast_{k,\nu}}
  = 0 \quad (\nu=1,\dots,d-1)
  \ .
\end{equation}
Consequently, the momentum $\vec{p}^\ast_k(t_f)$ has to be perpendicular to
the separatrix ${\cal S}(t_f)$ at the end point
$\vec{x}_\mathrm{sep}(\vec{s}^\ast_k, t_f)$. Furthermore, we need the
Hessian of the action as a function of the parameter vector~$\vec{s}$. Via
Eqs.~\eqref{3.10a} and \eqref{3.13} it can be expressed as
\begin{equation}
  \label{4.5}
  \begin{split}
  \frac{\partial^2 \phi_k(\vec{x}_\mathrm{sep}(\vec{s}^\ast_k,t_f),t_f)}
       {\partial s^\ast_{k,\nu} \partial s^\ast_{k,\nu'}}
  = {}&
  \frac{\partial \vec{x}_\mathrm{sep}(\vec{s}^\ast_k,t_f)}
       {\partial s^\ast_{k,\nu}}
  \cdot
  \mat{G}^{\ast}_{k}(t_{f})
  \,
  \frac{\partial \vec{x}_\mathrm{sep}(\vec{s}^\ast_k,t_f)}
       {\partial s^\ast_{k,\nu'}} \\
  & +
  \vec{p}^\ast_k(t_f) \cdot
  \frac{\partial^2 \vec{x}_\mathrm{sep}(\vec{s}^\ast_k,t_f)}
       {\partial s^\ast_{k,\nu} \, \partial s^\ast_{k,\nu'}}
  \quad (\nu,\nu'=1,\dots,d-1)
  \ .
  \end{split}
\end{equation}
Later on, we shall evaluate the last equation within a
linearization around the unstable periodic orbit of our
deterministic dynamics. In that case, the second term on the r.h.s
of Eq.~\eqref{4.5} vanishes. The determinant of the Hessian is thus
given by the first term, which can be written as (cf. Appendix
\ref{appendixa}):
\begin{equation}
  \label{4.5a}
  \det
  \left(
  \frac{\partial \vec{x}_\mathrm{sep}(\vec{s}^\ast_k,t_f)}
       {\partial s^\ast_{k,\nu}}
  \cdot
  \mat{G}^{\ast}_{k}(t_{f})
  \,
  \frac{\partial \vec{x}_\mathrm{sep}(\vec{s}^\ast_k,t_f)}
       {\partial s^\ast_{k,\nu'}}
  \right)_{\nu,\nu'=1,\dots, d-1}\!\!
  =
  g(\vec{s}^\ast_k, t_f)\,
  \vec{n}(\vec{s}^\ast_k, t_f)
  \cdot
  \mat{G}^\ast_k(t_f)^{-1}
  \,
  \vec{n}(\vec{s}^\ast_k, t_f)\,
  \det \mat{G}^\ast_k(t_f) \,
\end{equation}
Using the fact that the momentum $\vec{p}^\ast_{k}(t_f)$ is
parallel to the normal vector $\vec{n}(\vec{s}^\ast_k, t_f)$ (cf.\
Eq.~\eqref{4.4}), we then obtain for the rate~\eqref{4.2} the
important intermediate result
\begin{equation}
  \label{4.6}
  \Gamma(t_f) =
  \frac{1}{\sqrt{2^{d+1} \pi \epsilon}}
  \sum_k
  \frac{
    \vec{p}^\ast_k(t_f) \cdot
    \mat{D} \,
    \vec{p}^\ast_k(t_f)\,
    e^{-\phi_{k}(\vec{x}_\mathrm{sep}(\vec{s}^\ast_k,t_f),t_f)/\epsilon}
  }
  {\sqrt{
      \vec{p}^\ast_k(t_f)
      \cdot
      \mat{G}^\ast_k(t_f)^{-1}
      \,
      \vec{p}^\ast_k(t_f)\,
      \det \mat{Q}^{\ast}_{k}(t_{f})\,
      \det \mat{G}^{\ast}_{k}(t_{f})
  }
  }\,
  [1+\Ord(\epsilon)]\ .
\end{equation}

\subsection{The action minimizing paths}

\label{paths}

To proceed further in the evaluation of the rate formula~\eqref{4.6}, it is
necessary to gain more insight about the nature of the action minimizing paths
$\vec{x}^\ast_k(t)$, whose dynamics is governed by the Hamiltonian
equations~\eqref{3.8} supplemented by the boundary conditions~\eqref{3.9}. In
view of our choice for the initial conditions (cf.\ discussion after
Eq.~\eqref{2.8}) and our result~\eqref{4.6}, the latter assume the form
\begin{equation}
  \label{5.1}
  \vec{x}^{\ast}_{k}(t_{0}) = \vec{x}_s(t_0)
  \quad,\quad
  \vec{x}^{\ast}_{k}(t_{f}) = \vec{x}_\mathrm{sep}(\vec{s}^\ast_k, t_f)\ ,
\end{equation}
where the parameter values $\vec{s}^\ast_k$ are restricted by the
relations~\eqref{4.4}.

For given values of $t_0$ and $t_f$, in the generic case only one of the
solutions $\vec{x}^\ast_k(t)$ of this boundary problem will represent a global
minimum of the action~\eqref{3.10}. Denoting this path for the moment by
$\vec{x}^\ast_{\bar{k}}(t)$, it is clear that for large time differences
$t_f-t_0$ owing to the form of the action~\eqref{3.4} and \eqref{3.5} this
path $\vec{x}^\ast_{\bar{k}}(t)$ will spent most of its time near a
deterministic trajectory, \textit{i.e.},
$\dot{\vec{x}}^\ast_{\bar{k}}(t)\approx \vec{F}(\vec{x}^\ast_{\bar{k}}(t),t)$.
With regard to the boundary conditions~\eqref{5.1}, this means that after its
start at time $t_0$ the path will closely follow the stable periodic orbit
$\vec{x}_s(t)$ for some time.  Subsequently it switches over into the vicinity
of the separatrix ${\cal S}(t)$, thereby accumulating the main part of its action,
to remain there until its end at time $t_f$.  With respect to the position of
the end point on the separatrix, if we assume that ultimately all
deterministic trajectories on the separatrix converge to the unstable periodic
orbit $\vec{x}_u(t)$, the same will also hold true for the path
$\vec{x}^\ast_{\bar{k}}(t)$, which will thus end nearer and nearer to
$\vec{x}_u(t)$ the more time it is able to spend in a close vicinity of the separatrix.
Since the duration of the sojourns near the periodic orbits is long
compared to that of the transition in between, the path
$\vec{x}^\ast_{\bar{k}}(t)$ is often called an ``instanton'' in the
literature.

If we consider now the limiting case $t_0\to-\infty$ and $t_f\to\infty$ (in
the following abbreviated as $t_f-t_0\to\infty$), there exists a well defined
limit of $\vec{x}^\ast_{\bar{k}}(t)$, in the sense that this path follows ever
closer the periodic orbits, while retaining the shape of the intermediate
segment.  At the same time, its action $S[\vec{x}^\ast_{\bar{k}}(t)]$
converges from above to a finite value. We observe that in the limit
$t_f-t_0\to\infty$ owing to the time-periodicity of the force field
$\vec{F}(\vec{x},t)$ the action $S[\vec{x}^\ast_{\bar{k}}(t+n\T)]$ is the same
for all integers $n$. In other words, the global minimum of the action becomes
countable infinitely degenerate. However, one can still safely assume that these minima are
well separated in the space of all the paths appearing in~\eqref{3.3},
provided that the driving is neither too weak, too slow nor too
fast. These limiting cases are thus not covered by our theory.  We remind the
reader of the situation in the static case, where the global minimum is also
degenerate. However, while the degeneracy there is continuous in time
(Goldstone mode), we are dealing here with a discrete degeneracy.

As a consequence of the fact that the minimizing paths $\vec{x}^\ast_k(t)$ remain well
separated, our rate-formula (\ref{4.6}) becomes asymptotically exact
for any (arbitrary but fixed) finite values of the driving amplitude and
period as the noise strength $\epsilon$ tends to zero.  Apart from this fact
that {\em in the limit} $\epsilon\to 0$ the $\Ord (\epsilon)$ correction in
the saddle point approximation (\ref{3.7}) and thus in (\ref{4.6}) vanishes, a
more detailed quantitative statement seems difficult.  On the other hand, for
a given (small) noise strength $\epsilon$, we have to exclude extremely small
driving amplitudes and extremely long or short driving periods since this
would lead us effectively back to the static (undriven) escape problem, which
requires a completely different treatment (especially of the (quasi-)
Goldstone mode \cite{sme99,wei79,car81,col75,wei81,wei87}) than in
(\ref{3.7}).  Put differently, in any of these three asymptotic regimes, the
error $\Ord (\epsilon)$ from (\ref{3.7},\ref{4.6}) becomes very large.

For the following, we introduce the symbol $\vec{x}^\ast_\mathrm{opt}(t)$ for
the limit of the path $\vec{x}^\ast_{\bar{k}}(t)$ for $t_f-t_0\to\infty$,
always keeping in mind that this path is only defined modulo time shifts by
integer multiples of the driving period $\T$. The corresponding action is
defined analogously by
\begin{equation}
  \label{5.2}
  \phi_\mathrm{opt} := S[\vec{x}^\ast_\mathrm{opt}(t)]\ .
\end{equation}
Furthermore, all other quantities related to
$\vec{x}^\ast_\mathrm{opt}(t)$ inherit the subscript ``opt'',
for instance $\vec{p}^\ast_\mathrm{opt}(t)$ to name only one.

\begin{figure}[t]
  \includegraphics[width=0.5\linewidth]{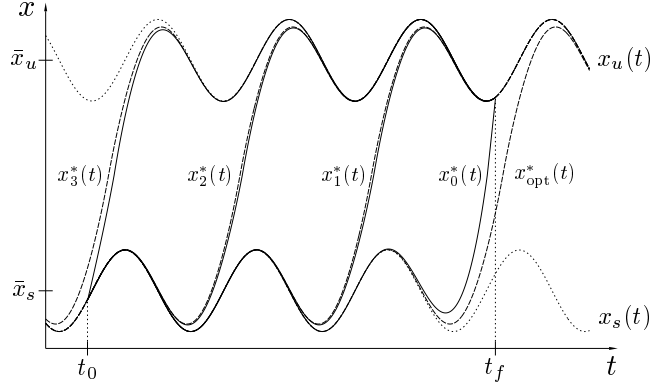}
  \caption{Action minimizing paths $x^\ast_k(t)$,
    $k=0,\dots,K(t_f,t_0)=3$ (solid) and corresponding master paths
    $x^\ast_\mathrm{opt}(t+k\T)$ (dashed) between stable and unstable
    periodic orbits (dotted). Depicted is the one-dimensional,
    overdamped case $d=1$ with an additively, harmonically driven
    piecewise parabolic potential from Ref.~\cite{leh00a,leh00b}.}
  \label{fig:paths}
\end{figure}

Coming back to the case of finite times $t_0$ and $t_f$, we expect that as
precursors of the limit $t_f-t_0\to\infty$ there exist besides the global
minimum $\vec{x}^\ast_{\bar{k}}(t)$ additional, relative minima
$\vec{x}^\ast_{k}(t)$ with an only slightly larger action. After a suitable
relabeling, each of them closely resembles an appropriately shifted
``master path'' $\vec{x}^\ast_\mathrm{opt}(t+k\T)$ (see Fig.~\ref{fig:paths}).
For finite $t_f-t_0$, we have a finite
number of the order $t_f-t_0/\T$ of such paths.  Thus, again
without restriction of generality, we can assume that the sum in the rate
formula~\eqref{4.6} runs from $0$ to a maximal value $K(t_f,t_0)$:
\begin{equation}
  \label{5.3}
  0\le k \le K(t_f,t_0) = {\cal O}((t_f-t_0)/\T)\ .
\end{equation}
Especially, the path $\vec{x}^\ast_0(t)$ is the one that stays as long as possible
in the vicinity of the stable periodic orbit $\vec{x}_s(t)$ and starts with its
transition towards the separatrix at the latest possible moment.

\subsection{Linearization scheme around  periodic orbits}

\label{linearization}

The discussion in the last section has shown that the paths
$\vec{x}^\ast_k(t)$ spend most of their time near the periodic orbits
$\vec{x}_{s,u}(t)$ of the deterministic dynamics. Thus, we may gain further
insight in their behavior
if we linearize the time-dependent force field around these orbits: 
\begin{equation}
  \label{5.4}
  \vec{F}(\vec{x}, t) \approx
  \vec{F}(\vec{x}_{s,u}(t),t) +
  \bnabla \vec{F}(\vec{x}_{s,u}(t),t)
  (\vec{x} - \vec{x}_{s,u}(t))\ .
\end{equation}
Within this approximation, the Hamiltonian equations~\eqref{3.8}
assume the form
\begin{equation}
  \label{5.5}
  \begin{split}
    \Delta\dot{\vec{x}}^\ast_k(t) & =
    \bnabla\vec{F}(\vec{x}_{s,u}(t),t) \, \Delta \vec{x}^\ast_k(t)
    + 2 \, \mat{D} \,\vec{p}^\ast_k(t) \\
    \dot{\vec{p}}^\ast_k(t) & =
    - \left(\bnabla \vec{F}(\vec{x}_{s,u}(t), t)\right)^{T}
    \vec{p}^{\ast}_{k}(t)\ ,
  \end{split}
\end{equation}
where we have introduced the deviations from the periodic orbits,
\begin{equation}
  \label{5.6}
  \Delta \vec{x}^\ast_k(t) := \vec{x}^\ast_k(t) - \vec{x}_{s,u}(t)\ .
\end{equation}

As we shall see later, for the evaluation of the rate
formula~\eqref{4.6}, one only needs the solution for the momentum
equation in the second line of Eq.~\eqref{5.5}.  According to
Floquet's theory \cite{flo83}, this solution can be obtained from the eigenvalues
and the left eigenvectors of the eigenvalue problem
\begin{equation}
  \label{5.7}
  \left[\frac{\d}{\d t} + \bnabla\vec{F}(\vec{x}_{s,u}(t),t)\right]
  \boldsymbol\Phi^\alpha_{s,u}(t)
   =
   \lambda^\alpha_{s,u}\,  \boldsymbol\Phi^\alpha_{s,u}(t)
   \ ,\
  \boldsymbol\Phi^\alpha_{s,u}(t+\T) = \boldsymbol\Phi^\alpha_{s,u}(t)
  \ ,
  \alpha=1,\dots, d
\end{equation}
together with the corresponding right eigenvectors
$\boldsymbol\Phi^{\dagger,\alpha}_{s,u}(t) $, \textit{i.e.}\ the
solutions of Eq.~\eqref{5.7} with
$\left(\bnabla\vec{F}(\vec{x}_{s,u}(t),t)\right)^T$. Upon proper
normalization of these eigenvectors at equal times, \textit{i.e.}
\begin{equation}
  \boldsymbol\Phi^{\dagger,\alpha}_{s,u}(t) \cdot
  \boldsymbol\Phi^{\beta}_{s,u}(t) =
  \delta_{\alpha\beta}\ ,
\end{equation}
we are able to write the solution for the momentum equation as
\begin{equation}
  \label{5.8}
  \vec{p}^\ast_k(t) =
  \sum_\alpha
  e^{-\lambda^\alpha_{s,u} (t-t_1)} \,
  \boldsymbol\Phi^{\alpha}_{s,u}(t_1)\cdot
  \vec{p}^\ast_k(t_1)\,
  \boldsymbol\Phi^{\dagger,\alpha}_{s,u}(t)
  \ ,
\end{equation}
where $t_1$ are arbitrary reference times for which the
linearization~\eqref{5.4} is valid. Since for the master path we have
$\Delta \vec{x}^\ast_\mathrm{opt}(t)\to 0$ for $t\to\pm\infty$, it
follows with Eq.~\eqref{5.5} that
$\lim_{t\to\pm\infty}\vec{p}^\ast_\mathrm{opt}(t)=0$. Consequently, we
must require $\mathop\mathrm{Re} \lambda^\alpha_u > 0$
($\mathop\mathrm{Re} \lambda^\alpha_s < 0$) for all $\alpha$ with
$\boldsymbol\Phi^{\alpha}_{u,s}(t_1)\cdot
\vec{p}^\ast_\mathrm{opt}(t_1)\ne 0$.  In the following, we denote
sums over this subset of eigenvectors by an apostrophe on the sum
sign.

With respect to the prefactor quantities $\mat{G}^\ast_k(t)$ and $\det
\mat{Q}^\ast_k(t)$, we first observe that within the linearization~\eqref{5.4}
the matrix Riccati equation~\eqref{3.14} assumes the form
\begin{equation}
  \label{5.9}
  \dot{\mat{G}}^{\ast}_{k}(t) =
  - 2\, \mat{G}^{\ast}_{k}(t)\, \mat{D}\, \mat{G}^{\ast}_{k}(t)
  - (\bnabla \vec{F}(\vec{x}_{s,u}(t),t))^T\,\mat{G}^{\ast}_{k}(t)
   -   \mat{G}^{\ast}_{k}(t)\,\bnabla \vec{F}(\vec{x}_{s,u}(t),t)
  \ .
\end{equation}
A transformation to the inverse matrix $\mat{G}^\ast_{k}(t)^{-1}$ yields the
linear matrix differential equation
\begin{equation}
  \label{5.10}
  \frac{\d}{\d t}
  \left[{\mat{G}}^{\ast}_{k}(t)^{-1}\right] =
  2\, \mat{D}
  + {\mat{G}}^{\ast}_{k}(t)^{-1} \, (\bnabla  \vec{F}(\vec{x}_{s,u}(t),t))^T
  + \bnabla \vec{F}(\vec{x}_{s,u}(t),t) \, {\mat{G}}^{\ast}_{k}(t)^{-1}
  \ ,
\end{equation}
where the initial condition is given by Eq.~\eqref{3.15} as
\begin{equation}
  \label{5.11}
  \mat{G}^\ast_{k}(t_0)^{-1} = \mat{0}.
\end{equation}
Furthermore, multiplication of Eq.~\eqref{5.9} with $\mat{G}^\ast_{k}(t)^{-1}$
shows that
\begin{equation}
  \label{5.12}
  \Tr \left[ \dot{\mat{G}}^\ast_k(t) \mat{G}^\ast_k(t)^{-1} \right]
  =
  -2 \Tr
  \left[
    (\bnabla  \vec{F}(\vec{x}_{s,u}(t),t))^T + \mat{D} \, \mat{G}^\ast_k(t)
  \right]\ .
\end{equation}
With the help of the linearized version of Eq.~\eqref{3.16} we then find for the
last two terms of the product appearing in the denominator of our rate
expression~\eqref{4.6} the result
\begin{equation}
  \label{5.13}
  \det \mat{G}^\ast_k(t)   \det \mat{Q}^\ast_k(t)
  = \mathrm{const.} =: \mu_{s,u}\ .
\end{equation}
Note that while $\mu_s$ is fixed for all paths $\vec{x}^\ast_k(t)$ by the
initial condition~\eqref{3.15} to the value $\mu_s = 2^{-d}$, $\mu_u$ depends
both on the index $k$ and on the explicit form of the time-dependent force
field $\vec{F}(\vec{x}, t)$. Later on, we only need $\mu_u$ for the master
path, which we denote by
\begin{equation}
  \label{5.13a}
  \mu_\opt :=
  \lim_{t\to\infty}
  \det \mat{G}^\ast_\mathrm{opt}(t)
  \det \mat{Q}^\ast_\mathrm{opt}(t) \ .
\end{equation}
The rest of the product in the denominator of \eqref{4.6} can be rewritten
by using the equation of motion~\eqref{5.10} for $\mat{G}^\ast_k(t)^{-1}$
together with the Hamiltonian equation~\eqref{5.5} for $\vec{p}^\ast_k(t)$.
This yields
\begin{equation}
  \label{5.14}
  \frac{\d}{\d t}
  \left[
    \vec{p}^\ast_k(t) \cdot
    \mat{G}^\ast_k(t)^{-1} \,
    \vec{p}^\ast_k(t)
  \right]
  =
  2 \, \vec{p}^\ast_k(t) \cdot \mat{D} \, \vec{p}^\ast_k(t)
\end{equation}
or equivalently in integrated form
\begin{equation}
  \label{5.15}
    \vec{p}^\ast_k(t) \cdot
    \mat{G}^\ast_k(t)^{-1} \,
    \vec{p}^\ast_k(t)
    =
    \vec{p}^\ast_k(t_1) \cdot
    \mat{G}^\ast_k(t_1)^{-1} \,
    \vec{p}^\ast_k(t_1)
    +
    2
    \int^t_{t_1} \!\!\d t'\,
    \vec{p}^\ast_k(t') \cdot \mat{D} \, \vec{p}^\ast_k(t')\ .
\end{equation}
For the master path $\vec{x}^\ast_\mathrm{opt}(t)$, the integral on the right
hand side of the last equation has to converge in the limit $t\to\infty$ in
order that the action $\phi_\mathrm{opt}$ assumes a finite value. Hence, the
quantity on the left hand side is also well defined in this limit and we
obtain for $t\to\infty$ after a subsequent renaming $t_u\to t$
\begin{equation}
  \label{5.16}
    \vec{p}^\ast_\mathrm{opt}(t) \cdot
    \mat{G}^\ast_\mathrm{opt}(t)^{-1} \,
    \vec{p}^\ast_\mathrm{opt}(t) =
    q_\mathrm{opt} -
   \vec{p}^\ast_\mathrm{opt}(t)\cdot
    \mat{A}_u(t)\,
    \vec{p}^\ast_\mathrm{opt}(t)
\end{equation}
Here we have exploited Eq.~\eqref{5.8} for the master path and
furthermore introduced the abbreviations
\begin{align}
  \label{5.17}
  q_\mathrm{opt} & :=
  \lim_{t\to\infty}
  \vec{p}^\ast_\mathrm{opt}(t) \cdot
  \mat{G}^\ast_\mathrm{opt}(t)^{-1} \,
  \vec{p}^\ast_\mathrm{opt}(t)\ ,\\
  \label{5.18}
  \mat{A}_u(t) & :=
  2
  \sideset{}{'}\sum_{\alpha\beta}
  \boldsymbol\Phi^\alpha_u(t)
  \boldsymbol\Phi^\beta_u(t)^T
  \int^\infty_t
  \d t'\,
  e^{-(\lambda^\alpha_u+\lambda^\beta_u)(t'-t)}
  \boldsymbol \Phi^{\dagger,\alpha}_u(t')\cdot
  \mat{D}\,
  \boldsymbol \Phi^{\dagger,\beta}_u(t')\ .
\end{align}
Note that owing to the time-periodicity of the Floquet solutions,
$\mat{A}_u(t)$ is also a periodic function of $t$:
\begin{equation}
  \label{5.19}
  \mat{A}_u(t+k\T) = \mat{A}_u(t) \quad \text{for all times $t$.}
\end{equation}

\subsection{Approximation in terms of the master path}

After having gained sufficient insight into the nature of the paths
$\vec{x}^\ast_k(t)$, we now come back to the evaluation of the rate
expression~\eqref{4.6}.  The main idea is to approximate all quantities
related to the paths $\vec{x}^\ast_k(t)$ in terms of the corresponding master
path, since, as discussed in Sect.~\ref{paths}, both resemble each other
closely.  However, while the boundary conditions for the master path are
$\vec{x}^\ast_\mathrm{opt}(t)-\vec{x}_s(t)\to 0$ for $t\to-\infty$ and
$\vec{x}^\ast_\mathrm{opt}(t)-\vec{x}_u(t)\to 0$ for $t\to\infty$, the paths
$\vec{x}^\ast_k(t)$ have to satisfy Eq.~\eqref{5.1}.  From the results of our
previous work~\cite{leh00b}, we know that while we can safely neglect these
deviations at the initial time $t_0$, the boundary at time
$t_f$ has to be treated more carefully.  First, we modify the upper
boundary condition for $\vec{x}^\ast_\mathrm{opt}(t)$ by requiring that
\begin{equation}
  \label{6.1}
  t_k := t_f + k\T
  \ ,\quad
  \vec{x}_k := \vec{x}^\ast_\mathrm{opt}(t_k)
\end{equation}
are the new end time and end point, respectively. In other words, we simply
truncate the path $\vec{x}^\ast_\mathrm{opt}(t+k\T)$ corresponding to
$\vec{x}^\ast_k(t)$ at the time $t_f$. Obviously, this new path still
satisfies the Hamiltonian equations~\eqref{3.8} and is thus an extremizing
path. The value of its action follows from the definitions~\eqref{3.10} and
\eqref{5.2} as
\begin{equation}
  \label{6.2}
  \phi_\mathrm{opt}(\vec{x}_k, t_k) :=
  \int_{-\infty}^{t_k} \!\d t\,\,
  \vec{p}^{\ast}_\mathrm{opt}(t)\cdot\mat{D}\,\vec{p}^{\ast}_\mathrm{opt}(t) =
  \phi_\mathrm{opt} -
  \int_{t_k}^\infty \!\d t\,\,
  \vec{p}^{\ast}_\mathrm{opt}(t)\cdot\mat{D}\,\vec{p}^{\ast}_\mathrm{opt}(t)\ .
\end{equation}

Next, we express the action~\eqref{3.10} of the path $\vec{x}^\ast_k(t)$ by
expanding the one belonging to the associated master path
$\vec{x}^\ast_\mathrm{opt}(t+k\T)$ in powers of the difference $\delta
\vec{x}^\ast_k(t_f) := \vec{x}^\ast_k(t_f) - \vec{x}^\ast_\mathrm{opt}(t_k)=
\Delta \vec{x}^\ast_k(t_f) - \Delta \vec{x}^\ast_\mathrm{opt}(t_k)$ between
their endpoints,
\begin{equation}
  \label{6.3}
  \phi_k(\vec{x}_\mathrm{sep}(\vec{s}^\ast_k,t_f),t_f) =
  \phi_\mathrm{opt}(\vec{x}_k, t_k) +
  \frac{\partial \phi_\mathrm{opt}(\vec{x}_k, t_k)}{\partial \vec{x}_k}
  \cdot
  \delta \vec{x}^\ast_k(t_f)
  + \cdots \ .
\end{equation}
The second term on the right hand side can be rewritten as
$\vec{p}^\ast_\mathrm{opt}(t_k)\cdot \Delta \vec{x}^\ast_k(t_f) -
\vec{p}^\ast_\mathrm{opt}(t_k)\cdot\Delta \vec{x}^\ast_\mathrm{opt}(t_k)$.
Here, the first scalar product is zero, as can been seen as follows: We know
that the endpoint $\vec{x}^\ast_k(t_f)$ lies one the separatrix ${\cal
  S}(t_f)$ and thus a deterministic solution $\vec{x}_\mathrm{det}(t)$
starting at time $t_k$ from $\vec{x}_\mathrm{det}(t_k) = \vec{x}^\ast_k(t_f)$
will for all times stay on ${\cal S}(t)$. Its
deviation from the unstable periodic orbit is denoted by
$\Delta \vec{x}_\mathrm{det}(t)$. Within our linearization~\eqref{5.4} one can
then readily verify with the help of the dynamical equation~\eqref{5.5} that $\d [
\vec{p}^\ast_\mathrm{opt}(t)\cdot\Delta \vec{x}_\mathrm{det}(t)]/\d t = 0$.
Hence, this scalar product is a constant of motion and we can infer that
\begin{equation}
  \label{6.4}
  \vec{p}^\ast_\mathrm{opt}(t_k) \cdot \Delta \vec{x}^\ast_k(t_f) =
  \vec{p}^\ast_\mathrm{opt}(t_k) \cdot \Delta \vec{x}_\mathrm{det}(t_k) =
  \lim_{t\to\infty} \left[
    \vec{p}^\ast_\mathrm{opt}(t) \cdot \Delta \vec{x}_\mathrm{det}(t)
  \right]
  = 0\ .
\end{equation}
Here, the last equality follows from the boundedness of
$\vec{x}_\mathrm{det}(t)$ near the saddle point (cf. the discussion below
Eq.~\eqref{2.2}) together with $\vec{p}^\ast_\mathrm{opt}(t)\to0$ for
$t\to\infty$.

The second scalar product, $\vec{p}^\ast_\mathrm{opt}(t_k)\cdot\Delta
\vec{x}^\ast_\mathrm{opt}(t_k)$, can be determined by a similar consideration:
The linearized Hamiltonian dynamics~\eqref{5.5} yields
\begin{equation}
    \label{6.5}
    \frac{\d}{\d t}
    \left[
      \vec{p}^\ast_\mathrm{opt}(t)\cdot
      \Delta \vec{x}^\ast_\mathrm{opt}(t)
    \right] =
    2 \,
    \vec{p}^\ast_\mathrm{opt}(t)\cdot
    \mat{D}\,
    \vec{p}^\ast_\mathrm{opt}(t)\ ,
\end{equation}
and since again this scalar product has to vanish in the limit $t\to\infty$,
we obtain
\begin{equation}
  \label{6.6}
  \vec{p}^\ast_\mathrm{opt}(t_k)\cdot\Delta
  \vec{x}^\ast_\mathrm{opt}(t_k) =
  - 2 \int_{t_k}^\infty \!\d t\,\,
  \vec{p}^{\ast}_\mathrm{opt}(t)\cdot\mat{D}\,\vec{p}^{\ast}_\mathrm{opt}(t)\ .
\end{equation}
Altogether, the approximation~\eqref{6.3} for the value of the action of the
path $\vec{x}^\ast_k(t)$ thus takes the form
\begin{equation}
  \label{6.7}
  \phi_k(\vec{x}_\mathrm{sep}(\vec{s}^\ast_k,t_f),t_f) =
   \phi_\mathrm{opt} +
  \int_{t_f}^\infty \!\d t\,\,
  \vec{p}^{\ast}_\mathrm{opt}(t+k\T)\cdot\mat{D}\,\vec{p}^{\ast}_\mathrm{opt}(t+k\T) +
   \cdots\ .
\end{equation}
The $k$-dependence of the exponential term in the rate expression~\eqref{4.6}
is thus reduced to a simple sum of exponential factors as given by the
time-dependence~\eqref{5.8} of the integrand in \eqref{6.7}. With respect to
the prefactor terms in the rate expression~\eqref{4.6}, we already now from
the results of our previous work \cite{leh00b} that up to corrections of the
order $\Ord(|\vec{p}^\ast_\mathrm{opt}(t_k)|^2)$ we can directly substitute
all quantities belonging to the path $\vec{x}^\ast_k(t)$ by those of the
corresponding master path $\vec{x}^\ast_\mathrm{opt}(t+k\T)$. Consistently, we
also neglect the terms on the right hand side of Eq.~\eqref{5.16}, which are
of the same order.

\subsection{Rate formula}

\label{rateresult}

After the discussion in the previous section, it is just a matter of collecting
everything in order to arrive at the central result of the present
work, namely the asymptotic value of the instantaneous escape rate,
\begin{equation}
  \label{7.1}
  \Gamma(t) \simeq
  \sqrt{\epsilon} \,
  \alpha_\mathrm{opt}\,
  e^{-\phi_\mathrm{opt}/\epsilon}
  \kappa_\mathrm{opt}(t,\epsilon)\ .
\end{equation}
Here, we have introduced the quantities
\begin{align}
\label{7.2}
  \alpha_\mathrm{opt} & :=
  \left[2^{d+1} \pi \, \T^2
    \lim_{t\to\infty}
      \vec{p}^\ast_\mathrm{opt}(t)
      \cdot
      \mat{G}^\ast_\mathrm{opt}(t)^{-1}
      \,
      \vec{p}^\ast_\mathrm{opt}(t)\,
      \det \mat{Q}^{\ast}_\mathrm{opt}(t)\,
      \det \mat{G}^{\ast}_\mathrm{opt}(t)
  \right]^{-\frac{1}{2}}\ ,\\
\label{7.3}
  \kappa_\mathrm{opt}(t,\epsilon) & :=
  \T \sum_{k=0}^{K(t,t_0)}
  \frac{\vec{p}^\ast_\mathrm{opt}(t+k\T)\cdot \mat{D}\,
        \vec{p}^\ast_\mathrm{opt}(t+k\T)}
        {\epsilon}\,
  \exp
  \left[
    -  \frac{1}{\epsilon}
    \int_{t}^\infty \!\d t'\,\,
    \vec{p}^{\ast}_\mathrm{opt}(t'+k\T)\cdot\mat{D}\,\vec{p}^{\ast}_\mathrm{opt}(t'+k\T)
  \right]\ .
\end{align}
With the help of Eqs.~(\ref{5.8}, \ref{5.18}) and the time-periodicity of the Floquet
solutions $\boldsymbol\Phi^\alpha_u(t)$ we can rewrite the last expression as
\begin{align}
  \label{7.4}
  \kappa_\mathrm{opt}(t,\epsilon) & =
  \T \sum_{k=0}^{K(t,t_0)}
  \frac{
    \vec{b}_\mathrm{opt}(t) \cdot
    \mat{B}_k(t)^T\,
    \mat{D}\,
    \mat{B}_k(t)\,
    \vec{b}_\mathrm{opt}(t)
  }{\epsilon} \,
  \exp
  \left\{
    -
    \frac{
    \vec{b}_\mathrm{opt}(t) \cdot
    \mat{B}_k(t)^T\,
    \mat{A}_u(t)\,
    \mat{B}_k(t)\,
    \vec{b}_\mathrm{opt}(t)
    }{2\,\epsilon}
  \right\}
  \ ,\\
  \mat{B}_k(t)& :=
  \sideset{}{'}\sum_{\alpha}
  e^{-\lambda^\alpha_u k \T}
  \boldsymbol\Phi^{\dagger,\alpha}_{u}(t) \,
  \boldsymbol\Phi^{\alpha}_{u}(t)^T =
  \mat{B}_k(t+\T) \ ,
  \label{7.5}\\
  \label{7.5aa}
  \vec{b}_\mathrm{opt}(t) & :=
  \lim_{\hat{t}\to\infty}
  \sideset{}{'}\sum_\alpha
  e^{-\lambda^\alpha_u(t-\hat{t})}\,
  \boldsymbol\Phi^{\alpha}_{u}(\hat{t})\cdot
  \vec{p}^\ast_\mathrm{opt}(\hat{t})\,
  \boldsymbol\Phi^{\dagger,\alpha}_{u}(t)\ .
\end{align}

Since $\mat{B}_k(t)$ in (\ref{7.5}) is a sum of terms that exponentially decrease with $k$,
there is a competition in the sum~\eqref{7.4} between a pre-exponential factor
which quickly decreases with $k$ and an exponential term increasing with $k$.
The main contribution to this sum thus comes from a few $k$ values around a
number $\hat{k}(t)$ that is for small noise strength $\epsilon$ much larger
than $0$ but at the same moment, for large enough $t_f-t_0$, still much
smaller than $K(t_f,t_0)$. Hence, up to an exponentially small error in
$\epsilon$, we can extend the summation range in Eq.~\eqref{7.4} and thus in
Eq.~\eqref{7.3} to all integers $k$. This allows us to identify two important
features of the time-dependence of $\kappa_\mathrm{opt}(t,\epsilon)$. To
prove the first one, the time-periodicity
\begin{equation}
  \label{7.5a}
  \kappa_\mathrm{opt}(t+\T,\epsilon) =
  \kappa_\mathrm{opt}(t,\epsilon)\ ,
\end{equation}
we merely have to shift the summation index in Eq.~\eqref{7.3} by $1$.  For
asymptotic times and small noise strengths, the instantaneous escape
rate~\eqref{7.1} is thus, as expected, a periodic function of the time $t$.  To
establish the second property, we use furthermore that the pre-exponential
term in Eq.~\eqref{7.3} is just the negative time-derivative of the expression
in the exponential. Again up to an exponentially small correction, we then
obtain
\begin{equation}
  \label{7.6}
  \frac{1}{\T}
  \int_t^{t+\T}\!\!
  \d t'\,
  \kappa_\mathrm{opt}(t',\epsilon)
  = 1 \ .
\end{equation}

Inserting the asymptotic rate expression~\eqref{7.1} into the
definition~\eqref{2.12} for the time-averaged rate and using the last
identity, we thus arrive at the second main result of the present
work, namely
\begin{equation}
  \label{7.7}
  \bar\Gamma \simeq
  \sqrt{\epsilon}\,
  \alpha_\mathrm{opt}\,
  e^{-\phi_\mathrm{opt}/\epsilon}\ .
\end{equation}
Hence, the noise-strength dependence of the time-averaged rate is of
the form of an Arrhenius-type, exponentially leading term times an
$\epsilon$-dependent prefactor.  Note that for a system in thermal
equilibrium \textit{i.e.}\ for a time-independent force field
$\vec{F}(\vec{x})$, the escape rate is given by an exponentially
leading Arrhenius factor, which contains the barrier against the
escape, multiplied by an $\epsilon$-independent prefactor, which
depends only on local properties of the force field at the barrier and
in the well \cite{lan69}. In comparison with this equilibrium rate
structure, we observe two crucial differences in (\ref{7.7}). First,
in the present, non-equilibrium situation, both the effective
potential barrier in the exponential and the prefactor depend in a
non-trivial way on global properties of the force field
$\vec{F}(\vec{x},t)$, and can thus in general only be determined by
means of a numerical computation. Second, the non-equilibrium case
exhibits an $\epsilon$-dependence of the prefactor.

\section{Explicit results for a metastable piecewise parabolic potential}
\label{example}

While, in general, one has to resort to numerical methods for the evaluation
of the rate expression~\eqref{7.1}, in the special case of a particle moving
in a one-dimensional piecewise parabolic potential with an additive sinusoidal
driving~\eqref{2.3} a complete analytical treatment is possible. In the
following, we work out the simplest such example with two parabolic pieces,
and compare the so obtained analytical predictions with numerical results from
a Monte-Carlo simulation of the stochastic dynamics~\eqref{2.3}.

Let us thus consider the force field deriving from the piecewise parabolic
potential of the form
\begin{equation}
  \begin{split}
  V(x\le 0) & = \frac{1}{2}\,m\,\omega_s^{2}\left[(x-\xbs)^{2} - \xbs^{2} \right] \\
  V(x\ge 0) & = \frac{1}{2}\,m\,\omega_u^{2}\left[(x-\xbu)^{2} - \xbu^{2} \right] \ ,
  \end{split}
  \label{8.1}
\end{equation}
where $\xbs<0$ and $\xbu>0$ denote the position of the potential minimum and
maximum, respectively, with corresponding curvatures
\begin{equation}
m \, \omega_{s}^2 >0 \quad \text{and}\quad m \, \omega_{u}^2 <0\ .
\end{equation} Note that within our notation $\omega_u$ is an imaginary number.
To ensure the continuity of the corresponding force field $F(x,t)$,
\begin{equation}
  \begin{split}
  F(x\le 0,t) & = - m\, \omega_s^{2} \, (x-\xbs) + A\, \sin(\Omega\, t) \\
  F(x\ge 0,t) & = - m\, \omega_u^{2} \, (x-\xbu) + A\, \sin(\Omega\, t) \ ,
  \end{split}
  \label{8.2}
\end{equation}
at the point $x=0$, we have to impose the additional restriction
$\omega_s^{2}\,\xbs = \omega_u^{2}\,\xbu$. Selecting as independent model
parameters $\omega_s^{2}$, $\omega_u^2$, and the static potential barrier
$\Delta V:=V(\xbu)-V(\xbs)$, the fixed points $x_{s,u}$ can be expressed
through
\begin{equation}
  \label{8.3}
  \omega_s^{2}\,\xbs =
  \omega_u^{2}\,\xbu =
  - \sqrt{\frac{2\, \Delta V}{m}\,
          \frac{\omega_s^{2}\, |\omega_u|^{2}}
               {\omega_s^{2}+|\omega_u|^{2}}}\ .
\end{equation}
If we require furthermore that the periodic orbits $x_{s,u}(t)$ do not cross
the point $x=0$, \textit{i.e.}
\begin{equation}
  \label{8.4}
  x_s(t) < 0 < x_u(t)
\end{equation}
for all times $t$, which is granted if and only if the conditions
\begin{equation}
  \label{8.5}
  A^{2} <
  m^{2}
  \left[
    \gamma^{2}\,\Omega^{2} + (\Omega^{2} - \omega_{s,u}^{2})^{2}
  \right]
  \xbsu^{2}
\end{equation}
are fulfilled for both the ``s'' and the ``u'' indices, we obtain
\begin{equation}
  \label{8.6}
  x_{s,u}(t) =
  \xbsu -
  \frac{A}{m}\,
  \frac{ \gamma\, \Omega \,\cos(\Omega\, t) + (\Omega^{2} - \omega_{s,u}^{2})\, \sin(\Omega \, t)}
       {\gamma^{2}\, \Omega^{2} + (\Omega^{2} - \omega_{s,u}^{2})^{2}}\ .
\end{equation}

For the determination of the master path $x^\ast_\mathrm{opt}(t)$, we use the
specific form~\eqref{2.4} of the force field and of the diffusion matrix to
express the Hamiltonian equations of motion~\eqref{3.8} in the form of the
second order differential equations
\begin{equation}
  \label{8.7}
  \begin{split}
    \Delta\ddot{x}^\ast_\mathrm{opt}(t) +
    \gamma \, \Delta \dot{x}^\ast_\mathrm{opt}(t)
    + \omega_{s,u}^2 \, \Delta x^\ast_\mathrm{opt}(t)
    & =
    2 \, \frac{\gamma}{m} \, p^\ast_{v,\mathrm{opt}}(t)\\
    \ddot{p}^\ast_{v,\mathrm{opt}}(t) -
    \gamma \, \dot{p}^\ast_{v,\mathrm{opt}}(t) +
    \omega_{s,u}^2 \, p^\ast_{v,\mathrm{opt}}(t) & = 0 \ .
  \end{split}
\end{equation}
Here, the index ``s'' (``u'') applies for all times $t$ where
$x^\ast_\mathrm{opt}(t)\le 0$ ($x^\ast_\mathrm{opt}(t)\ge 0$).
Furthermore, we have already set the strength $\delta$ of the
auxiliary noise term appearing in~\eqref{2.4} to $0$, since
Eq.~\eqref{8.7} and all the following relations remain well-defined for a
singular diffusion matrix $\mat{D}$.  The system~\eqref{8.7} of linear
equations can be readily solved, if we restrict ourselves to the case
where the master path $x^\ast_\mathrm{opt}(t)$ crosses the point $x=0$
exactly once, say at time $t_1$,
\begin{equation}
  \label{8.8}
  x^\ast_\mathrm{opt}(t) = 0 \quad \Longleftrightarrow \quad t=t_{1} \ .
\end{equation}
Notice that owing to the form of the Hamiltonian equations \eqref{8.7}
both $x^\ast_\mathrm{opt}(t)$ and $p^\ast_{v,\mathrm{opt}}(t)$ have to
be continuously differentiable at this time $t_1$.  For the following
considerations it is convenient to introduce the frequencies
\begin{equation}
  \label{8.9}
  \lambda^{\pm}_{s,u} =  -\frac{\gamma}{2} \pm
  \sqrt{\frac{\gamma^{2}}{4} - \omega_{s,u}^{2}}\ ,
\end{equation}
which are the Floquet eigenvalues from Eq.~\eqref{5.7}. We observe
that $\lambda^-_u<0$, such that the respective terms do not appear in
the sums~(\ref{5.18},\ref{7.5},\ref{7.5aa}). The corresponding Floquet
states are given by
\begin{equation}
 \boldsymbol\Phi^\pm_{s,u}(t) =
\frac{1}{\sqrt{\lambda^+_{s,u}-\lambda^-_{s,u}}}
\begin{pmatrix}1 \\ \lambda^\pm_{s,u}\end{pmatrix}
\quad\text{and}\quad
\boldsymbol\Phi^{\dagger,\pm}_{s,u}(t) =
\frac{1}{\sqrt{\lambda^+_{s,u}-\lambda^-_{s,u}}}
\begin{pmatrix} \lambda^\pm_{s,u}+\gamma \\ 1
\end{pmatrix}
\ .
\end{equation}
Hence, the matrices $\mat{B}_k(t)$ from Eq.~\eqref{7.5} take the form
\begin{equation}
  \mat{B}_k(t) =
  \frac{e^{-\lambda^+_u k \T}}{\lambda^+_u-\lambda^-_u}
  \begin{pmatrix}
    \lambda^+_u+\gamma & 1 \\
    \lambda^+_u(\lambda^+_u+\gamma) & \lambda^+_u
  \end{pmatrix}\ .
\end{equation}
Taking into account the boundary conditions $\Delta
x^\ast_\mathrm{opt}(t)\to 0$ and $p^\ast_{v,\mathrm{opt}}(t)\to 0$ for
$t\to\pm\infty$, we obtain for the solutions of Eq.~\eqref{8.7}
\begin{equation}
  \begin{split}
  \Delta x^\ast_\mathrm{opt}(t\le t_{1}) & =
  \frac{1}{\lambda^+_s-\lambda^-_s}
  \left\{
  \left[
    \frac{1}{m}\,p^\ast_{v,\mathrm{opt}}(t_{1})
    - \lambda^+_s\,
    \xs(t_{1})
  \right]
  e^{-\lambda^-_s \cdot (t-t_{1})}\, -
  \left[
    \frac{1}{m}\,p^\ast_{v,\mathrm{opt}}(t_{1}) -
    \lambda^-_s  \xs(t_{1})
  \right]
  e^{- \lambda^+_s\cdot(t-t_{1})}
  \right\}
  \\
  \Delta x^\ast_\mathrm{opt}(t\ge t_{1}) & =
  \frac{1}{\lambda^+_u}
  \left\{
  \left[
    \frac{1}{m}\,p^\ast_{v,\mathrm{opt}}(t_{1}) -
     \lambda^+_u \,\xu(t_{1})
  \right]
  e^{\lambda^-_u\cdot(t-t_{1})} \, -
  \frac{1}{m}\,p^\ast_{v,\mathrm{opt}}(t_{1}) \,
  e^{- \lambda^+_u\cdot(t-t_{1})}
  \right\}
  \\
  p^\ast_{v,\mathrm{opt}}(t\le t_{1}) & =
  \frac{1}{\lambda^+_s - \lambda^-_s}
  \left\{
  \left[
    -\lambda^-_s  p^\ast_{v,\mathrm{opt}}(t_{1}) +
    m\,\omega_s^{2}\, \xs(t_{1})
  \right]
  e^{ -\lambda^-_s \cdot(t-t_{1})} +
  \left[
    \lambda^+_s\,  p^\ast_{v,\mathrm{opt}}(t_{1}) -
    m\,\omega_s^{2}\, \xs(t_{1})
  \right]
  e^{- \lambda^+_s\cdot(t-t_{1})}
  \right\}
  \\
  p^\ast_{v,\mathrm{opt}}(t\ge t_{1}) & =
  p^\ast_{v,\mathrm{opt}}(t_{1})\,
  e^{-\lambda^+_u\cdot(t-t_{1})}\ .
\end{split}
\label{8.10}
\end{equation}

Next one can infer from the above mentioned requirement of a continuous first
derivative at time $t$ of both $x^\ast_\mathrm{opt}(t)$ and
$p^\ast_{v,\mathrm{opt}}(t)$ the two relations
\begin{align}
  p^\ast_{v,\mathrm{opt}}(t_{1}) & =  m\, \lambda^+_u
  \left[
    \xu(t_{1})
    +\frac{\dot{x}_s(t_{1})-\dot{x}_u(t_{1})}{\lambda^-_u}
  \right]
  \nonumber\\
  p^\ast_{v,\mathrm{opt}}(t_{1}) & = m \, \frac{\omega_s^{2}}{\omega_u^{2}}\,\lambda^+_u\,
  \xs(t_{1})
  \ .
  \label{8.11}
\end{align}
These relations can be used to fix the two so far unknown quantities $t_1$ and
$p^\ast_{v,\mathrm{opt}}(t_{1})$ in the above expressions:
A straightforward, but somewhat tedious calculation
yields
\begin{align}
  \tan (\Omega\, t_{1}) & = \frac{1}{\Omega}\, \frac{\big(\omega_s^{2}
    - \Omega^{2}\big)\lambda^+_u + \gamma\,\Omega^{2}}
  {\omega_s^{2}-\Omega^{2} - \gamma\,\lambda^+_u}
  \label{8.12}
  \\
  p^\ast_{v,\mathrm{opt}}(t_{1}) & =
   m\,\lambda^+_u\,\xbu
  + \omega_s^{2}\,\frac{\lambda^+_u}{\lambda^-_u}\,
  \frac{A}{\Omega}\,
  \frac{\cos(\Omega\,t_{1})}{\omega_s^{2}-\Omega^{2} - \gamma\,\lambda^+_u}\ .
  \label{8.13}
\end{align}
Obviously, Eq.~\eqref{8.12} has two solutions within each driving period
$\T$. We anticipate that only one corresponds to a minimum of the action, and
hence to the master path. Thus, we fix $t_1$ (up to the usual degeneracy under
$t\to t+\T$) by requiring additionally that
\begin{equation}
  \label{8.14}
  \frac{A}{\Omega}\,
  \frac{\cos(\Omega \,t_{1})}
       {\omega_s^{2}-\Omega^{2} - \gamma\,\lambda^+_u}
       > 0 \ .
\end{equation}
Combining \eqref{8.12}--\eqref{8.14} we arrive at
\begin{equation}
  p^\ast_{v,\mathrm{opt}}(t_{1})  =
  m \, \lambda^+_u\, \xbu
  + \frac{|A|\, \omega_s^{2}\, |\omega_u|^{2}}
         {\lambda^-_u\,\nu^{4}} > 0\ ,
  \label{8.15}
\end{equation}
where we have introduced the frequency $\nu$ via
\begin{equation}
\nu^{4} :=
  \left[
    \bigg(
      \gamma^{2} \Omega^{2}
      +
      \Big(
        \Omega^{2} - \omega_s^{2}
      \Big)^{2}
    \bigg)
    \bigg(
      \omega_u^{4} +
      \Omega^{2}\,
      {\lambda^-_u}^{2}
    \bigg)
  \right]^{1/2}\ .
  \label{8.16}
\end{equation}
To conclude the discussion of the master path, we note that in order to check
the consistency of the solution~\eqref{8.10} with the condition~\eqref{8.8} it
is necessary to solve a transcendental equation, which has to be done
numerically. Qualitatively, one expects a break-down of Eq.~\eqref{8.8} in the
deterministically underdamped regime, \textit{i.e.}\ $\gamma/2 < \omega_s$,
where the path $(x_\mathrm{opt}(t), v_\mathrm{opt}(t))$ in phase space leaves
the vicinity of the stable periodic orbit by way of a spiraling orbit which
crosses the line $x=0$ several times.

Inserting
$p^\ast_{v,\mathrm{opt}}(t)$ from Eq.~\eqref{8.10} and
(\ref{8.11},\ref{8.15}) into the definition (\ref{3.10},\ref{5.2})
yields the action of the master path:
\begin{equation}
  \label{8.17}
  \phi_\mathrm{opt} =
  \Delta V
  \left[
    1 - \left|
          \frac{A^{2}\, \omega_s^{2}\, \omega_u^{2}\, (\omega_s^{2}+ |\omega_u|^{2})}
               { 2\,m\,\Delta V\,\nu^{8}}
        \right|^{1/2}
  \right]^{2}\ .
\end{equation}
The periodic driving thus leads to an ``effective potential barrier''
$\phi_\mathrm{opt}$ which is smaller than the static barrier $\Delta
V$ to which it reduces in the limits $A\to0$ or $\Omega\to\infty$.
While $\phi_\mathrm{opt}$ is monotonically decreasing with increasing
driving amplitude~$A$, the dependency on the driving
frequency~$\Omega$ is more complicated. In particular one observes
resonance behavior near the frequency $\omega_s$ of the bottom well.
Had we chosen the opposite inequality in Eq.~\eqref{8.14}, a plus
instead of the minus sign would have appeared in the Eq.~\eqref{8.17}.
Thus, the condition~\eqref{8.14} does indeed single out the minimum of
the action and thus the desired solution for the master path
$x^\ast_\mathrm{opt}(t)$. Using the result for
$p^\ast_{v,\mathrm{opt}}(t_{1})$ from Eq.~\eqref{8.15}, we obtain the
following explicit expression for $\vec{b}_\mathrm{opt}(t)$ from
Eq.~\eqref{7.5aa}:
\begin{equation}
  \vec{b}_\mathrm{opt}(t) =
  \begin{pmatrix}
    \lambda^+_u + \gamma\\1
  \end{pmatrix}
  p^\ast_{v,\mathrm{opt}}(t_{1})\,
  e^{-\lambda^+_u(t-t_1)}
\end{equation}

Finally, we turn to the determination of the prefactor quantity $\alpha_\mathrm{opt}$.
We first note that in the present case the linear differential
equation~\eqref{5.10} does not only hold approximatively but is exact for
all times $t\ne t_1$. Owing to $t_0\to-\infty$, the solution has approached
its stationary value, and we obtain
\begin{equation}
  \label{8.18}
  \mat{G}^\ast_\mathrm{opt}(t < t_{1}) \equiv
  \begin{pmatrix}
    m\, \omega_s^{2} & 0\\
    0             & m
  \end{pmatrix}\
\end{equation}
and hence with Eq.~\eqref{3.15}
\begin{equation}
  \label{8.19}
  \det \mat{Q}^\ast_\mathrm{opt}(t<t_{1}) = \frac{1}{4\, m\, \omega_s^{2}}\ .
\end{equation}
At this point, we make use of the fact that $\det
\mat{Q}^\ast_\mathrm{opt}(t)$ is continuous at time $t_1$ according to Eq.~\eqref{3.16},
while $\mat{G}^\ast_\mathrm{opt}(t)$ jumps due to Eq.~\eqref{3.14}. With
$\partial^2 F(x,t)/\partial x^2 = m \left(\omega_s^{2}-\omega_u^{2}\right) \delta(x)$ one thus can
infer that
\begin{equation}
  \label{8.20}
  \mat{G}^\ast_\mathrm{opt}(t_{1} + 0)
  =
  \begin{pmatrix}
   \left(\omega_s^{2}+ |\omega_u|^{2}\right)
   \frac{m\,\dot{x}^\ast_\mathrm{opt}(t_{1}) - p^\ast_{v,\mathrm{opt}}(t_{1})}{\dot{x}^\ast_\mathrm{opt}(t_{1})}- |\omega_u|^{2}&
    0\\
    0 & m
  \end{pmatrix}\ .
\end{equation}
This yields together with Eq.~\eqref{5.13}, which again is exact for the
piecewise parabolic potential, the intermediate result
\begin{equation}
  \label{8.21}
  \lim_{t\to\infty} \det \mat{G}^\ast_\mathrm{opt}(t)\, \det \mat{Q}^\ast_\mathrm{opt}(t) =
  \frac{1}{4\,m\, \omega_s^{2}}
  \left[
    \left(\omega_s^{2}+ |\omega_u|^{2}\right)
    \frac{m\,\dot{x}^\ast_\mathrm{opt}(t_{1}) - p^\ast_{v,\mathrm{opt}}(t_{1})}{\dot{x}^\ast_\mathrm{opt}(t_{1})}
    - m|\omega_u|^{2}
  \right]\ .
\end{equation}
Applying two times Eq.~\eqref{5.15}, one time for $t=t_{0}\to-\infty$
using the initial condition~\eqref{5.11} and a second time for
$t\to\infty$, yields after some algebra another necessary quantity, namely
\begin{multline}
  \label{8.22}
  \lim_{t\to\infty}
    \vec{p}^\ast_\mathrm{opt}(t)\cdot\mat{G}^\ast_\mathrm{opt}(t)^{-1}\,\vec{p}^\ast_\mathrm{opt}(t)
  = \\\quad
  p^\ast_{x,\mathrm{opt}}(t_{1})^{2}
  \left\{
    \left[
      \left(\omega_s^{2} + |\omega_u|^{2} \right)
      \frac{m\,\dot{x}^\ast_\mathrm{opt}(t_{1}) - p^\ast_{v,\mathrm{opt}}(t_{1})}{\dot{x}^\ast_\mathrm{opt}(t_{1})}
      -m|\omega_u|^{2}
    \right]^{-1}\!\!
    -\frac{1}{m\,\omega_s^{2}}
  \right\}
  + 2 \,\phi_\mathrm{opt}\ .
\end{multline}
Inserting these expressions into the definition~\eqref{7.2} of
$\alpha_\mathrm{opt}$ and using Eqs.~\eqref{3.8},
\eqref{8.15} and \eqref{8.17} we arrive at
\begin{equation}
  \label{8.23}
  \alpha_\mathrm{opt} =
  \left[
    4\pi\,\T^{2} \,
    \frac{m\,\dot{x}^\ast_\mathrm{opt}(t_{1}) -
          p^\ast_{v,\mathrm{opt}}(t_{1})}
         {m\,\dot{x}^\ast_\mathrm{opt}(t_{1})}\,
    \phi_\mathrm{opt}
  \right]^{-1/2}\ .
\end{equation}
Again, our choice for the sign in Eq.~\eqref{8.14} is justified \textit{a
  posteriori}, since it guarantees that
$m\,\dot{x}^\ast_\mathrm{opt}(t_{1})-p^\ast_{v,\mathrm{opt}}(t_{1})$ is a
positive quantity.  Eventually, together with Eqs.~\eqref{8.3},
\eqref{8.6}, \eqref{8.12}, \eqref{8.13} and \eqref{8.15}, the
last expression can be rewritten as
\begin{equation}
  \label{8.24}
  \alpha_\mathrm{opt} =
  \left[
    \frac{|A|
          \left[
            \Omega^{2}{\lambda^-_u}^{2} - \omega_s^{2}\,  |\omega_u|^{2}
          \right]
          +
          \sqrt{\frac{2\,m\,\Delta V \, \nu^{8}}{\omega_s^{-2}+|\omega_u|^{-2}}}
         }
         {16\pi^{3}\,|A| {\lambda^-_u}^{2}\phi_\mathrm{opt}}
  \right]^{1/2}\ .
\end{equation}
We remark that the structure of the results \eqref{8.17} and
\eqref{8.24} closely resembles those in the overdamped case, as obtained in
Ref.~\cite{leh00a,leh00b}. In particular, this limiting case $m\to0$
is correctly reproduced by \eqref{8.17} and \eqref{8.24}.

We have checked our above analytical predictions by comparing them
with results from Monte-Carlo simulations of the stochastic
dynamics~\eqref{2.3}, which yield the mean time $\overline{t_{\mathrm{exit}}}$
necessary for an exit out of the driven well. The time-averaged rate is
related to this quantity via \cite{Rei99}
\begin{equation}
  \label{8.25}
  \bar\Gamma = \left(\,\overline{t_{\mathrm{exit}}}\,\right)^{-1}\ .
\end{equation}
Since the necessary simulation times diverge exponentially
with decreasing noise strength $\epsilon$ (cf. (\ref{7.7}) and (\ref{8.24})),
the numerical determination of the mean exit
time becomes impossible for extremely small  $\epsilon$.
For similar reasons, the numerical determination of the
time-resolved rates $\Gamma(t)$ is ruled out.

\begin{figure}
  \begin{center}
    \includegraphics[width=0.5\linewidth]{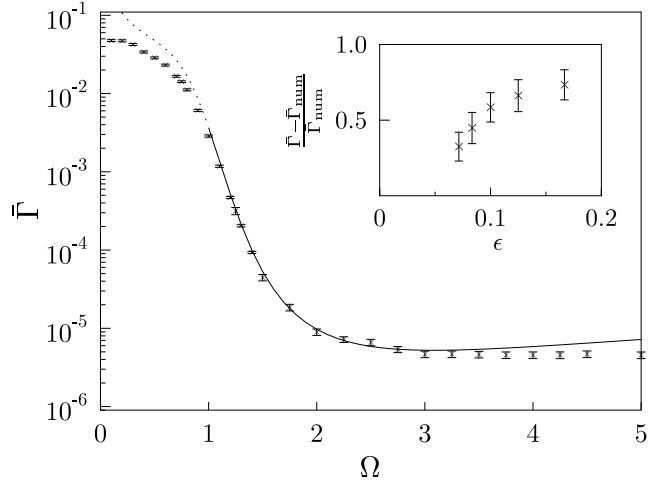}
    \caption{Time-averaged rate $\bar\Gamma$ \textit{vs.} driving frequency $\Omega$ for
      $\xbs=\omega_u^{2}=-1$, $\xbu=\omega_s^{2}=m=\eta=\Delta V=A=1,
      \epsilon=0.1$ (dimensionless units).  Crosses: Inverse mean
      first exit time from simulations of Eq.~\eqref{2.3} (error
      bars: mean square deviation of mean value). Solid (dotted) line:
      Analytical prediction~\eqref{7.7}
      (condition~\eqref{8.8} not fulfilled). Inset:
      Convergence behavior of $\bar\Gamma$ for $\epsilon\to 0$ for
      fixed driving frequency $\Omega=5$ (all other parameters as in
      main panel).  }
    \label{fig:rate_omega}
  \end{center}
\end{figure}

\begin{figure}
  \begin{center}
    \includegraphics[width=0.5\linewidth]{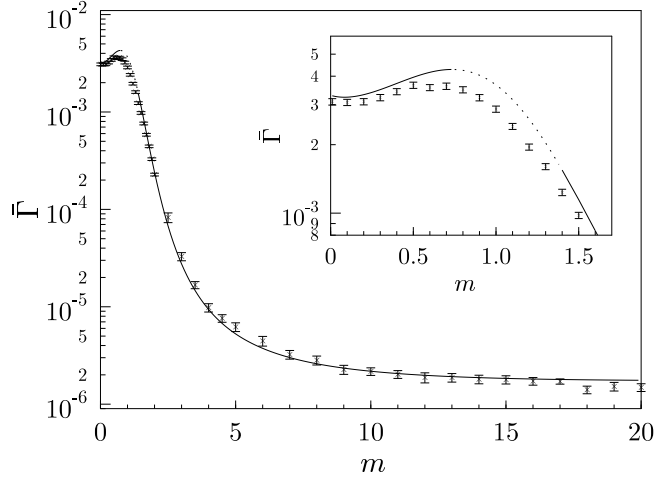}
    \caption{Time-averaged rate $\bar\Gamma$ \textit{vs.} particle
      mass $m$ for $\xbs=m\,\omega_u^{2}=-1$,
      $\xbu=m\,\omega_s^{2}=\eta=\Delta V=A=\Omega=1, \epsilon=0.1$
      (dimensionless units).  Crosses: Inverse mean first exit time
      from simulations of Eq.~\eqref{2.3} (error bars: mean square
      deviation of mean value). Solid (dotted) line: Analytical
      prediction~\eqref{7.7} (condition~\eqref{8.8} not
      fulfilled). Inset: Magnification of small $m$ regime. Note that
      the deterministic resonance lies at $m=\sqrt{1/2}\approx 0.707$.}
    \label{fig:rate_m}
  \end{center}
\end{figure}

\begin{figure}
  \begin{center}
    \includegraphics[width=0.5\linewidth]{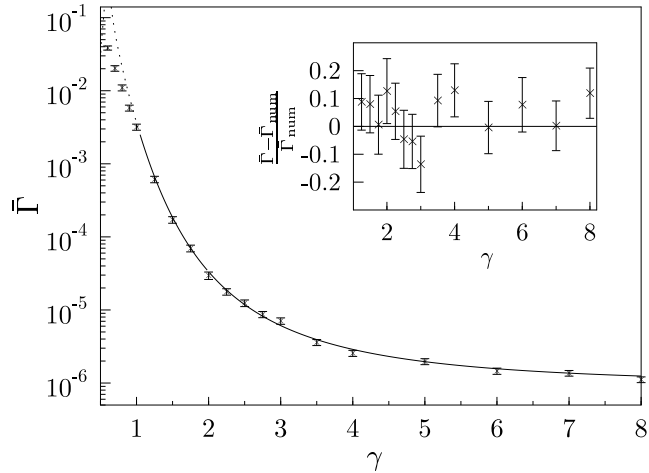}
    \caption{Time-averaged rate $\bar\Gamma$ \textit{vs.} damping coefficient
      $\gamma=\eta/m$ for $\xbs=\,\omega_u^{2}=-1$,
      $\xbu=\,\omega_s^{2}=m=\Delta V=A=\Omega=1, \epsilon=0.1$
      (dimensionless units).  Crosses: Inverse mean first exit time
      from simulations of Eq.~\eqref{2.3} (error bars: mean square
      deviation of mean value). Solid (dotted) line: Analytical
      prediction~\eqref{7.7} (condition~\eqref{8.8} not
      fulfilled). Inset: Relative difference between analytical
      prediction and numerical result as a function of $\gamma$.}
    \label{fig:rate_gamma}
  \end{center}
\end{figure}

Figures \ref{fig:rate_omega}--\ref{fig:rate_gamma} depict the time-averaged
rate $\bar\Gamma$ as a function of various system parameters.
Parameter regions for which condition~\eqref{8.8} is not
fulfilled are indicated, which in particular applies for the case of
weak damping $\eta$ and the vicinity of the deterministic resonance at
$\Omega^2=\omega_s^2 - \gamma^2/2$, as discussed above.
Outside of these regions, already for a
noise strength of $\epsilon=1$, the agreement between theory and simulation is
very good (compare also inset of Fig.~\ref{fig:rate_gamma}).

Furthermore, as
one can see in Fig.~\ref{fig:rate_omega}, the theory breaks down both for
small and large driving frequencies $\Omega$, if the noise strength $\epsilon$
is kept fixed.  However, for $\epsilon\to 0$, we again observe the convergence
of numerically determined rate towards the theoretical approximation (cf. inset of
Fig.~\ref{fig:rate_omega}).  This behavior is in full accordance with our
predictions from Sect.~\ref{paths}.

\section{Conclusions}
\label{conclusion}

With the present work, we have generalized earlier results for the
noise-activated escape in time-periodically driven one-dimensional
overdamped systems to an arbitrary number of dimensions. The basic
idea of the path-integral approach put forward is that it is
necessary to sum over all local, nearby minima of the relevant
action that contribute to the escape rate. For asymptotically weak
noise, these minima are well separated in the space of all paths,
and the contribution of each of them can be obtained by a standard
saddle-point approximation of the path-integral. This behavior is
in strong contrast to the undriven situation, where the occurrence
of a (quasi-) Goldstone mode requires a more sophisticated
treatment of the associated prefactor.

Our central results express the time-instantaneous rate~\eqref{7.1}
and the time-averaged rate~\eqref{7.7} in terms of quantities
belonging to a master path, which has in general to be obtained
numerically as the solution of a minimization problem.  However, the
structure of the result already reveals two noteworthy, salient
differences compared to the static case: both the exponentially
leading Arrhenius factor and the prefactor not only depend on the
relative barrier height and the local properties of the potential
inside the well and at the barrier region, respectively, but in
addition these two quantities depend as well sensitively on the {\em
  global behavior} of the time-dependent metastable potential.
Additionally, a $\sqrt{\epsilon}$-dependence of the rate prefactor
appears as a consequence of the non-equilibrium situation. Finally, we
have been able to derive closed analytical rate expressions for a
distinctive case, namely the sinusoidally driven Kramers problem with
a metastable potential formed by two parabolas.

We also like to point out here that the treatment of driven escape in
meta-stable potential landscapes by use of path integral methods is
close in spirit to prominent work by Jozef T.~Devreese wherein he
pioneered the challenging problem of polarons in magnetic fields
\cite{Dev85,Dev96}: in both cases the effective Hamiltonian involves a
vector-potential like contribution which is linear in the canonical
momentum variable, see Eq.~(\ref{Ham}).  Then, as pointed out
repeatedly by Jozef, the standard variational principle based on the
Feynman-Jensen inequality no longer applies \cite{Dev92,Dev96}, but
requires instead an appropriate modification \cite{Dev92}.

\begin{acknowledgments}
  This work has been supported by the Studienstiftung des deutschen
  Volkes~(J.L.), the Graduiertenkolleg GRK-283, the
  Sonderforschungsbereich SFB-486 (J.L.,P.H.), and the
  Sonderforschungsbereich SFB-613 (P.R.). One of us (P.H.) likes to
  acknowledge many elucidating and stimulating scientific discussions
  with Jozef T. Devreese who, as we all know, is still very active not
  only within his beloved physics of polarons but increasingly also
  within the timely area of physics on the nanoscale. Jozef is still
  young enough to appreciate and to contribute great science!
\end{acknowledgments}

\appendix*

\section{}
\label{appendixa}

In this appendix, we shall prove Eq.~\eqref{4.5a}. Let us first abbreviate the
$d\times d-1$ matrix $\mat{S}$ composed of the vectors $\partial
\vec{x}_\mathrm{sep}(\vec{s}^\ast_k, t_f)/\partial s^\ast_{k,\nu}$,
$\nu=1,\dots,d-1$ as columns. For notational brevity, we suppress here and in
the following the asterix, all indices $k$, and the time arguments. Indicating
furthermore with $\mat{S}^i$, $i=1,\dots,d$, the $d-1\times d-1$ matrix, which
consists of $\mat{S}$ without row $i$ and with $\mat{G}^i_j$, $i,j=1,\dots,d$,
the matrix without row $i$ and column $j$, we can apply a generalized version
of the determinant multiplication theorem \cite{linearalgebra} to the left hand side of
Eq.~\eqref{4.5a}, yielding
\begin{equation}
  \det (\mat{S}\, \mat{G} \, \mat{S} ) =
  \sum_{ij}
  \det \mat{S}^j\,
  \det \mat{G}^i_j\,
  \det \mat{S}^i\ .
\end{equation}
Using a standard relation between the cofactors and the inverse of
a matrix \cite{linearalgebra}
\begin{equation}
  (-1)^{i+j} \det \mat{G}^i_j\, = (\mat{G}^{-1})_{ji}\, \det\mat{G}\ ,
\end{equation}
we obtain the intermediate result
\begin{equation}
  \det (\mat{S}\, \mat{G} \, \mat{S} ) =
  \vec{u}\cdot
  \mat{G}^{-1}\,
  \vec{u}\,
  \det \mat{G}\ .
\end{equation}
Here, we have introduced the $d$-dimensional vector
$u_i:=(-1)^{i+d}\det\mat{S}^i$, $i=1,\dots,d$. A comparison with
Eq.~\eqref{4.5a} shows that the proof is complete, if we are able to
demonstrate that $\vec{u} = \sqrt{g}\, \vec{n}$. To this end, we define
another auxiliary quantity, namely the $d\times d$ matrix
$\mat{M}:=(\mat{S}|\vec{n})$.  Since the normal vector $\vec{n}$ of the
separatrix $\cal S$ is orthogonal to all columns of $\mat{S}$, the determinant
of $\mat{M}$ is given by (cf. also Eq.~\eqref{4.3})
\begin{equation}
  \det\mat{M} = \sqrt{\det \mat{M}^T \mat{M}} =
  \sqrt{\det\mat{S}^T\mat{S}} = \sqrt{g}
\end{equation}
For the same reason, one has $(\mat{M}^{-1})_{di} = \mat{M}_{id}= n_i$, $i=1,\dots,d$, and
thus
\begin{equation}
  u_i = (-1)^{i+d}\,  \det \mat{M}^i_d = \det\mat{M} \,
  (\mat{M}^{-1})_{di} = \sqrt{g} \, n_i\ .
\end{equation}


\end{document}